\documentclass[11pt]{article} 
\pdfoutput=1
\usepackage{graphicx}
\usepackage{subcaption}
\usepackage{amssymb}
\usepackage{amsmath}
\usepackage[left]{lineno}
\usepackage{subcaption} 
\usepackage{booktabs}
\usepackage{multirow}
\usepackage[title]{appendix}
\usepackage{lineno}
\usepackage{relsize}
\usepackage{physics} 
\usepackage{siunitx} 
\usepackage{enumitem}
\setlist[itemize]{itemsep=0pt,topsep=\baselineskip} 
\usepackage{pgfplots}
\usepackage{pgfplotstable}
\usepackage{tikz,pgfplots}
\usepackage{hyperref}
\usepackage{xcolor}
\pgfplotsset{compat=1.14}
\newsavebox{\foobox}

\newcommand\T{\rule{0pt}{2.3ex}}       
\newcommand\B{\rule[-1.2ex]{0pt}{0pt}} 

\usepackage{url}
\usepackage{jheppub} 
\begin{document}

\pagenumbering{arabic}
\vspace*{-2.5\baselineskip}
\centerline{\LARGE EUROPEAN ORGANIZATION FOR NUCLEAR RESEARCH}
\vspace{15mm}
\begin{flushright}
CERN-EP-2025-167 \\
23 July 2025 \\
--- \\
Revised version: \\
20 October 2025 \\
\end{flushright}
\vspace{5mm}

\begin{center}
\Large{ 
\bf Searches for hidden sectors using \boldmath{$K^+\to\pi^+X$} decays
\\
\vspace{5mm}
}
\begin{NoHyper}
\renewcommand{\thefootnote}{\fnsymbol{footnote}}
The NA62 Collaboration
\footnote[1]{
Corresponding authors: F.~Brizioli, M.~Raggi, J.~Swallow, \\
email: francesco.brizioli@cern.ch, mauro.raggi@cern.ch, joel.christopher.swallow@cern.ch
}
\end{NoHyper}

\end{center}
\vspace{5mm}

\begin{abstract}

\centerline{\textbf{Abstract}}

\vspace{5mm}

Results from the study of the rare decays $K^+\to\pi^+\nu\bar{\nu}$, $K^{+}\rightarrow\pi^{+}\mu^{+}\mu^{-}$ and $K^{+}\rightarrow\pi^{+}\gamma\gamma$ at the NA62 experiment at CERN are interpreted in terms of improved limits for $\mathcal{B}(K^+\to\pi^+X)$ and coupling parameters of hidden-sector models, where $X$ is a mediator.
World-leading limits are achieved for dark photon, dark scalar and axion-like particle models.

\end{abstract}
\vspace{10mm}

\begin{center}
\em{Accepted for publication in JHEP}
\end{center}

\clearpage



\section{Introduction}
\label{sec:Introduction}

The NA62 experiment, described in detail in~\cite{NA62DetectorPaper}, is a fixed-target experiment at the CERN SPS, designed to perform a stringent test of the Standard Model (SM) by measuring the ultra-rare \mbox{$K^+\to\pi^+\nu\bar{\nu}$} decay, which offers sensitivity to physics beyond the SM (BSM) up to a mass scale of $\mathcal{O}(100\,\text{TeV})$.
Within the SM, this decay is highly suppressed and precisely calculated.
Using tree-level measurements of the CKM matrix elements as external inputs, the SM branching ratio is predicted to be $\mathcal{B}(K^{+}\rightarrow\pi^{+}\nu\bar{\nu})=(8.4\pm1.0)\times10^{-11}$~\cite{Buras:2015qea}, while using only meson mixing processes to eliminate the dependence on $|V_{cb}|$, the branching ratio is predicted to be $(8.60\pm0.42)\times10^{-11}$~\cite{Buras:2022wpw}. Using a full CKM parameter fit, a value of $(7.86\pm0.61)\times10^{-11}$ is predicted~\cite{DAmbrosio:2022kvb}.
The NA62 experiment reported the measurement 
$\mathcal{B}(K^{+}\rightarrow\pi^{+}\nu\bar{\nu})= (13.0^{+3.3}_{-3.0})\times10^{-11}$ 
using data collected in 2016--2022~\cite{Pnn2016paper,Pnn2017paper,PnnRun1Paper,NA62Pnn2122}.
Additionally, studies of the rare decays 
$K^{+}\rightarrow\pi^{+}\mu^{+}\mu^{-}$~\cite{NA62Kpimumu}
and
$K^{+}\rightarrow\pi^{+}\gamma\gamma$~\cite{NA62:2023olg}
were performed using 2017--2018 data.
In this context, BSM contributions, though suppressed by small coupling constants, may still lead to significant enhancement of the decay rate with respect to the SM expectation via the process $K^+ \to \pi^+ X$, where $X$ is a feebly-interacting BSM particle~\cite{Goudzovski:2022vbt}. 

In the following, the $K^{+}\rightarrow\pi^{+}\nu\bar{\nu}$ study performed using the 2016--2022 dataset~\cite{NA62Pnn2122} is interpreted as a search for the $K^{+}\rightarrow\pi^{+}X$ decay.
Additionally, the $K^{+}\rightarrow\pi^{+}\mu^{+}\mu^{-}$ decay analysis from the 2017--2018 dataset~\cite{NA62Kpimumu} is interpreted as a search for the prompt decay chain $K^{+}\rightarrow\pi^{+}X,\,X\rightarrow\mu^{+}\mu^{-}$.
For completeness, the previously published results related to the $\pi^{0}\rightarrow\text{invisible}$ decay search~\cite{NA62pi0inv} and the study of $K^{+}\rightarrow\pi^{+}\gamma\gamma$ decays~\cite{NA62:2023olg} are discussed.
Constraints are set on benchmark hidden-sector models as limits of $\mathcal{B}(K^{+}\rightarrow\pi^{+}X)$, where:
\begin{itemize}
    \item $X$ is {\em{invisible}} ($X_{\rm inv}$), meaning it is not detected experimentally: it may decay to dark matter particles (or neutrinos), or to visible SM particles but with a sufficiently long lifetime to escape the detector;
    \item $X$ decays to SM {\em{visible}} particles which are detected.
\end{itemize}
Relevant new physics benchmark models~\cite{Beacham:2019nyx} are summarised in table~\ref{tab:BCtable}, with additional details given in section~\ref{sec:DarkSectorConstraints}.
Constraints are set on the BC2, BC4, BC10 and BC11 models of~\cite{Beacham:2019nyx}, in which the mediator is a vector, scalar or pseudoscalar particle. 
In addition, minimally modified BC4-inv and BC10-inv scenarios are considered, which are equivalent to BC4 and BC10 except that the mediator only decays to hidden-sector particles.
Other models where the hidden-sector particle is a heavy neutral lepton ($N$) are investigated at NA62 in dedicated searches for $K^{+}\rightarrow\ell^{+}N$ ($\ell=\mu,e$)~\cite{NA62:2020mcv,NA62:2021bji} and $\pi^{+}\rightarrow e^{+}N$ decays~\cite{NA62:2025csa}.
Hidden-sector searches at NA62 with kaon decays are complementary to those performed in beam dump mode~\cite{NA62:2023qyn,NA62:2023nhs,NA62DumpMode}.

\begin{table}[ht] 
    \centering
    \caption{Summary of new physics benchmark models 
    relevant to NA62 searches for decays of the form $K^{+}\rightarrow \pi^{+}X$, where $X$ is a vector, scalar or pseudoscalar particle. 
    The last column lists the experimental signature(s) of the $X$ particle considered in each case.
    }
    \vspace{8pt}
    \begin{tabular}{|l|l|l|l|l|} 
    \hline
    Benchmark & BSM particle ($X$) & Type & Coupling to SM & Search \T\B \\
    \hline
    \hline
    BC1 & dark photon ($A^{\prime}$) & vector & $\varepsilon$ & $\mu^{+}\mu^{-}$ \T\B \\
    BC2 & dark photon ($A^{\prime}$) & vector & $\varepsilon$ & invisible \T\B \\
    \hline
    BC4 & dark scalar ($S$) & scalar & $\theta$ & invisible, $\mu^{+}\mu^{-}$ \T\B \\
    BC4-inv & dark scalar ($S$) & scalar & $\theta$ & invisible \T\B \\
    \hline
    BC10 & axion-like particle ($a$) & pseudoscalar & $C_{ff}$ (to fermions) & invisible, $\mu^{+}\mu^{-}$ \T\B \\
    BC10-inv & axion-like particle ($a$) & pseudoscalar & $C_{ff}$ (to fermions) & invisible \T\B \\
    BC11 & axion-like particle ($a$) & pseudoscalar & $C_{GG}$ (to gluons) & invisible, $\gamma\gamma$ \T\B \\
    \hline
    \end{tabular} 
    \label{tab:BCtable}
\end{table}
\newpage

\section{Searches for \boldmath{$K^{+}\rightarrow\pi^{+}X$} with invisible $X$}
\label{sec:NA62Measurements_inv}

\subsection{Interpretation of \boldmath{$K^{+}\rightarrow\pi^{+}\nu\bar{\nu}$} results}
The measurement of the $K^{+}\rightarrow\pi^{+}\nu\bar{\nu}$ decay rate by the NA62 experiment is described in detail in~\cite{NA62Pnn2122}.
It relies on precise momentum and time measurements of the incident $K^{+}$ and downstream $\pi^{+}$, charged particle identification and hermetic veto of other particles produced in abundant $K^{+}$ decay modes, 
using a variety of techniques and detectors specially designed for this purpose~\cite{NA62DetectorPaper}.
The resulting analysis allows reaching sensitivity to  branching ratios as low as $10^{-11}$ to $10^{-12}$.
 
Because the $\nu\bar{\nu}$ pair is not detected, the signal signature of the SM process is identical to that of the $K^{+}\rightarrow\pi^{+}X_{\rm inv}$ decay.
A peak search is performed in the $K^{+}\rightarrow\pi^{+}\nu\bar{\nu}$ signal regions of the $m_{\rm miss}^{2} = (P_{K}-P_{\pi})^{2}$ distribution, where $P_K$ ($P_\pi$) is the reconstructed four-momentum of the kaon (pion) in the initial (final) state.
The SM $K^{+}\rightarrow\pi^{+}\nu\bar{\nu}$ background is evaluated using $\mathcal{B}(K^{+}\rightarrow\pi^{+}\nu\bar{\nu})=8.4\times10^{-11}$.
Applying the procedure described in~\cite{NA62KpiXPaper}, 
model-independent upper limits for $\mathcal{B}(K^{+}\rightarrow\pi^{+}X)$ at $90\,\%$ confidence level (CL) were established in the mass ranges $0$--$110\,\text{MeV}/c^{2}$ and $150$--$260\,\text{MeV}/c^{2}$
using the 2016--2018 dataset
which includes $4.3\times10^{12}$ 
collected $K^{+}$ decays~\cite{PnnRun1Paper}.

\begin{figure}[tb]
\centering
\subfloat[]{\includegraphics[width=0.49\linewidth]{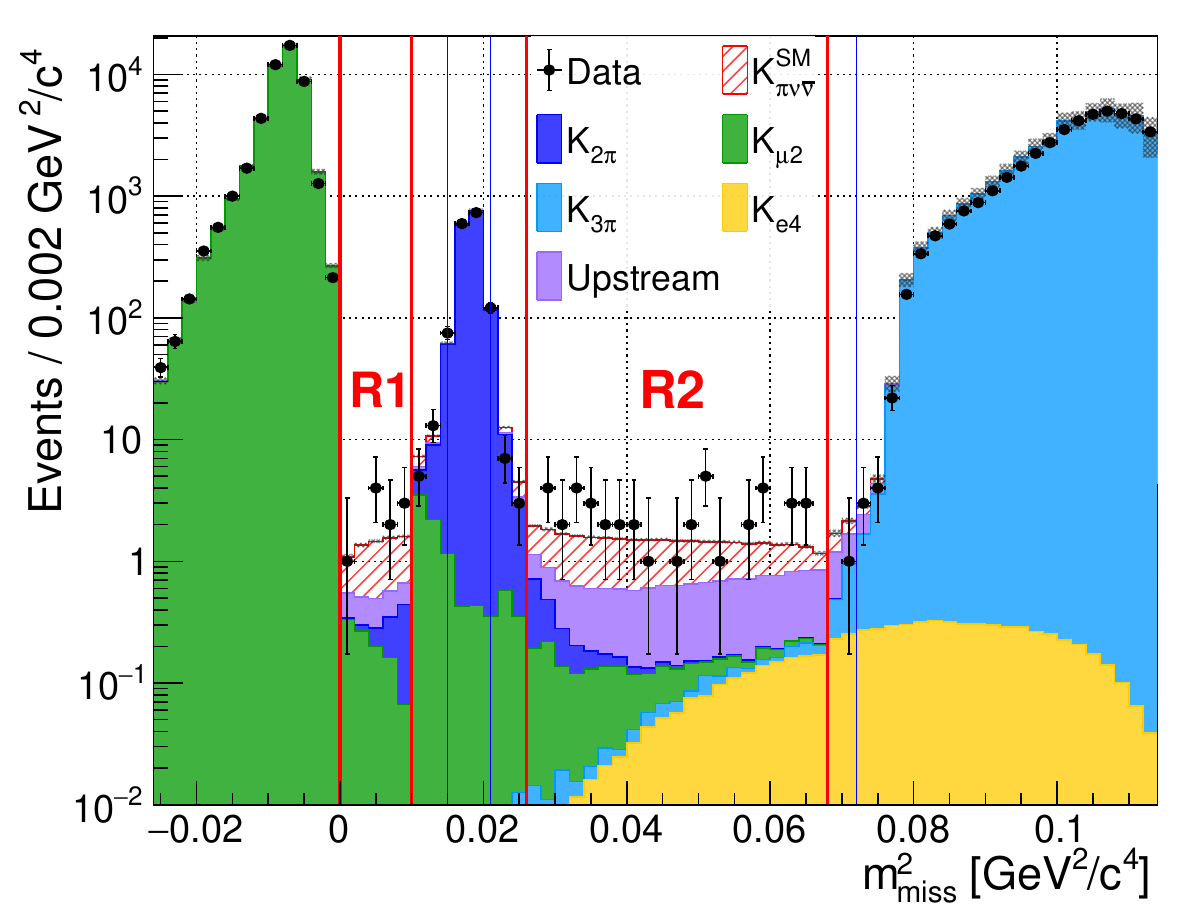}}\
\subfloat[]{\includegraphics[width=0.49\linewidth]{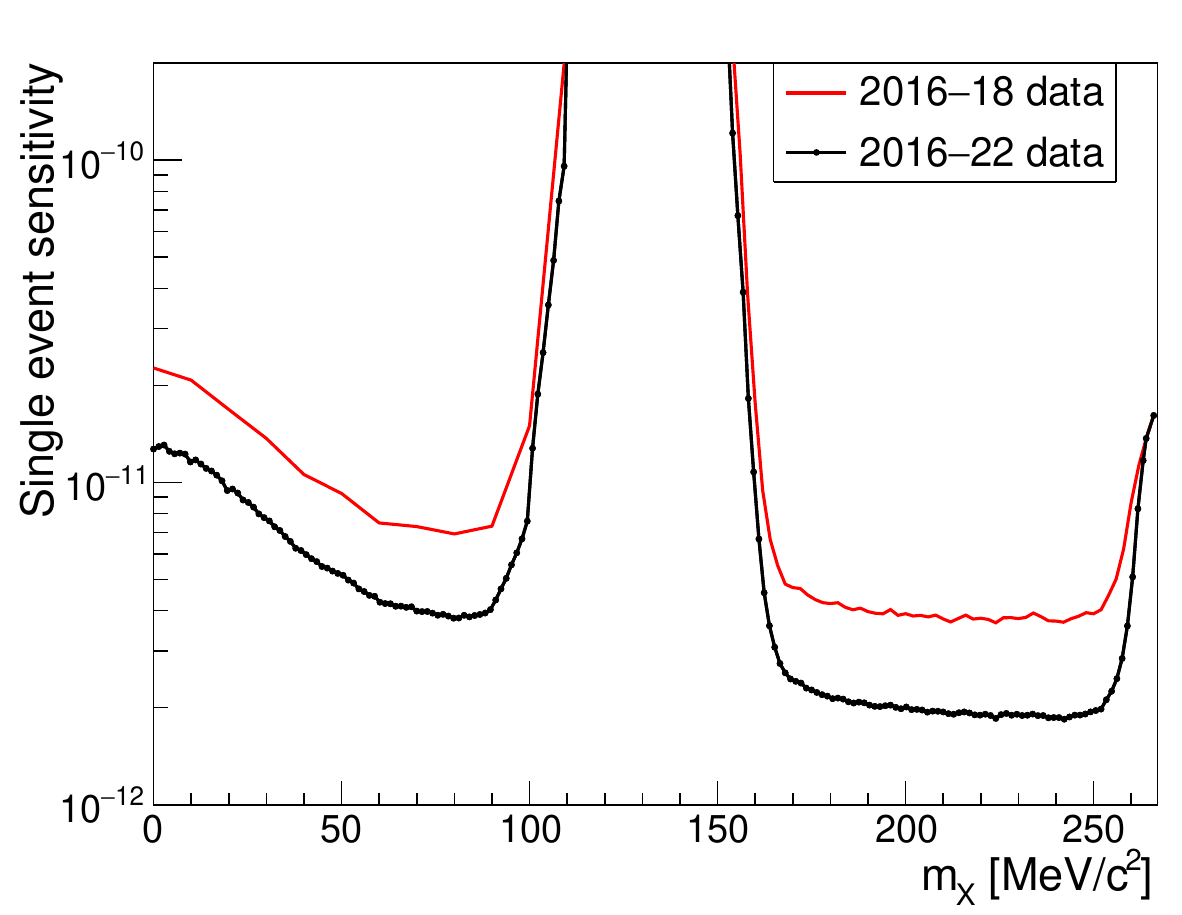}}\
\caption{
(a): Combined $m_{\rm miss}^{2}$ distributions for the 2016--2022 dataset, including the SM $K^{+}\rightarrow\pi^{+}\nu\bar{\nu}$ decay using 
$\mathcal{B}(K^{+}\rightarrow\pi^{+}\nu\bar{\nu})=8.4\times10^{-11}$~\cite{Buras:2015qea}, 
and background expectations.
(b): SES of the $K^{+}\rightarrow\pi^{+}X_{\rm inv}$ decay search as a function of $m_X$.
The SES obtained with the analysis of the 2016--2018 dataset~\cite{PnnRun1Paper} is also shown for comparison.
}
\label{fig:KpiX_1622_mm2_and_SES}
\end{figure}

\begin{figure}[tb]
\centering
\subfloat[]{\includegraphics[width=0.49\linewidth]{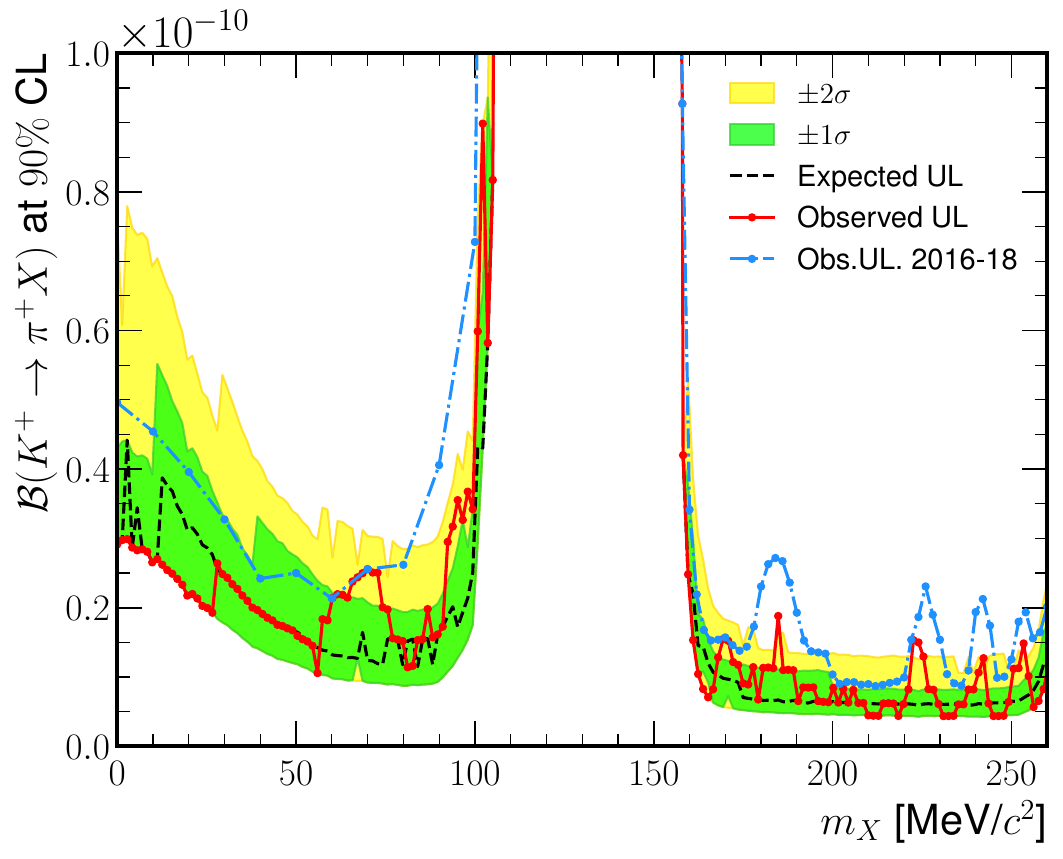}}\
\subfloat[]{\includegraphics[width=0.49\linewidth]{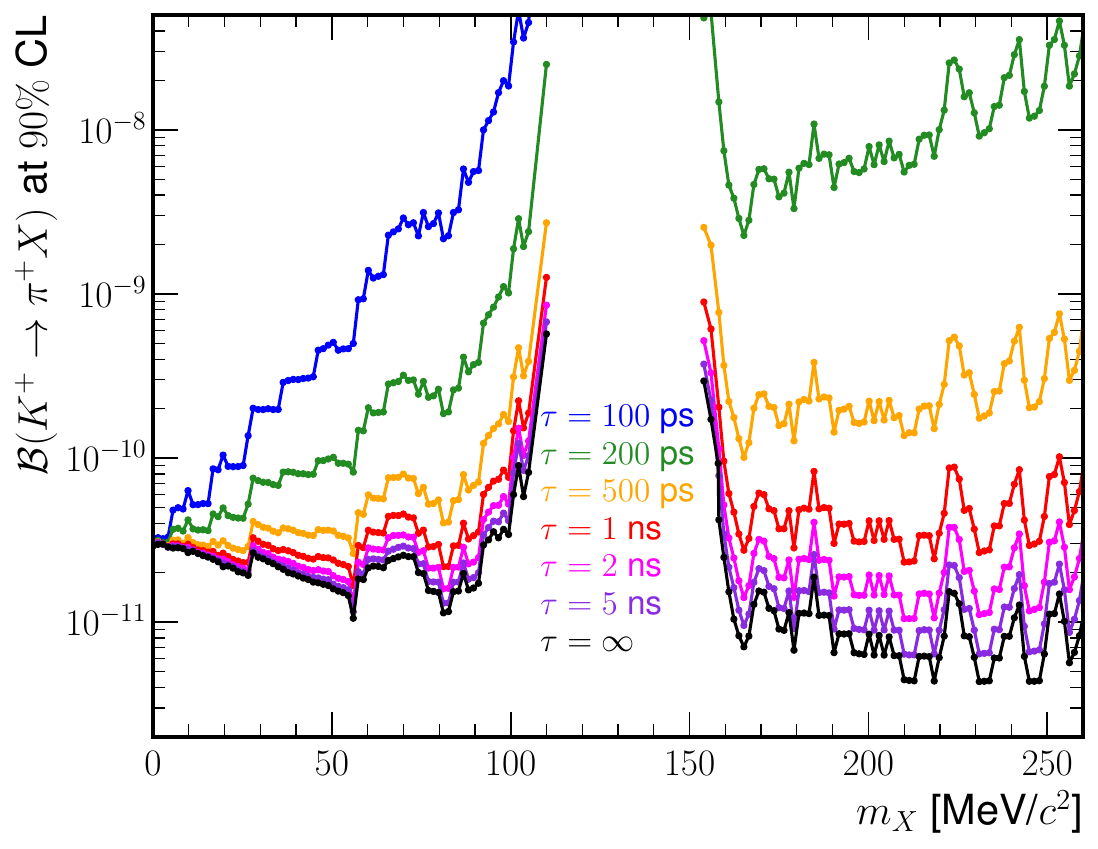}}\
\caption{
Model-independent constraints on $\mathcal{B}(K^+ \to \pi^+ X)$ as function of $m_X$, from interpretation of the 2016--2022 measurement of the $K^{+}\rightarrow\pi^{+}\nu\bar{\nu}$ decay.
(a): Expected and observed upper limits at $90\,\%$ CL for $\mathcal{B}(K^+ \to \pi^+ X)$ as function of $m_X$ for invisible $X$, with comparison to observed upper limits from 2016--2018 data~\cite{PnnRun1Paper}.
(b): Observed upper limits at $90\,\%$ CL of $\mathcal{B}(K^+ \to \pi^+ X)$ as function of $m_X$, for several $\tau_X$ hypotheses, assuming $X$ decays to visible SM particles.
}
\label{fig:KpiX_1622_limits}
\end{figure}

Here the search is extended to include the 2021--2022 dataset
which consists of $2.9 \times 10^{12}$ collected $K^+$ decays~\cite{NA62Pnn2122}.
The $m_{\rm miss}^{2}$ spectrum of expected and observed events satisfying the $K^{+}\rightarrow\pi^{+}\nu\bar{\nu}$ signal selection in the 2016--2022 dataset is shown in figure~\ref{fig:KpiX_1622_mm2_and_SES}-a. 
A scan is performed, searching for peaks in this spectrum, considering $m_{X}$ hypotheses separated by $1.4\,\text{MeV}/c^{2}$, which is less than the corresponding resolution for any mass hypothesis~\cite{NA62KpiXPaper}.
For each mass hypothesis a search window is defined, centered on $m_{X}^{2}$ and with a width of $3 \, \sigma_{m_{\text{miss}}^{2}}$
, where $\sigma_{m_{\text{miss}}^{2}}$ is the $m_{\rm miss}^{2}$ resolution
(established with simulations) which varies from $1.2\times10^{-3}\,\text{GeV}^2/c^{4}$ at $m_{X}=0$ to $0.7\times10^{-3}\,\text{GeV}^2/c^{4}$ at $m_{X} = 260\,\text{MeV}/c^{2}$~\cite{NA62KpiXPaper}.
In each search window the expected number of background events is calculated using the $m_{\rm miss}^{2}$ spectrum shown in figure~\ref{fig:KpiX_1622_mm2_and_SES}-a, and the number of observed events is counted.
An upper limit for the number of $K^{+}\rightarrow\pi^{+}X$ events in this search window, $N_{UL}^{\pi X}$, is evaluated using the $\text{CL}_{\text{S}}$ method~\cite{Read:2002hq}.
A model-independent limit of $\mathcal{B}(K^{+}\rightarrow\pi^{+}X) = \mathcal{B}_{\pi X}$, in the scenario where $X$ is invisible, is established according to 
\begin{equation}
    UL(\mathcal{B}_{\pi X}(m_{X})) 
        = N_{UL}^{\pi X}(m_{X}) \cdot \mathcal{B}_{SES}^{\pi X}(m_{X}) \,\,.
    \label{eqn:BrULs}
\end{equation}
Here $\mathcal{B}_{SES}^{\pi X}(m_{X})$ is the single event sensitivity (SES) for the $K^{+}\rightarrow\pi^{+}X$ search evaluated at the $m_X$ value, shown in figure~\ref{fig:KpiX_1622_mm2_and_SES}-b, given by 
\begin{equation}
    \mathcal{B}_{SES}^{\pi X}(m_{X}) = \mathcal{B}_{SES}^{\pi\nu\bar{\nu}}\frac{A_{\pi\nu\bar{\nu}}}{A_{\pi X}(m_{X})} \,\,,
    \label{eqn:Acc_SES}
\end{equation}
where $\mathcal{B}_{SES}^{\pi\nu\bar{\nu}}$ and $A_{\pi\nu\bar{\nu}}$ are the SES and acceptance for the SM $K^{+}\rightarrow\pi^{+}\nu\bar{\nu}$ decay~\cite{Pnn2016paper,Pnn2017paper,PnnRun1Paper,NA62Pnn2122}, and $A_{\pi X}(m_{X})$ is the acceptance of a $K^{+}\rightarrow\pi^{+}X$ decay, which depends on $m_{X}$.
The inclusion of the 2021--2022 data leads to an improvement in the SES by approximately a factor of $2$ (figure~\ref{fig:KpiX_1622_mm2_and_SES}-b). 
However the background, in particular from the SM $K^{+}\rightarrow\pi^{+}\nu\bar{
\nu}$ decay, 
becomes significant, with up to $2$ events expected per search window.
This restricts the improvement in the upper limit to a factor of $1$--$3$ 
depending on the mass hypothesis, as shown in figure~\ref{fig:KpiX_1622_limits}-a.
The effect of the SM $\mathcal{B}(K^{+}\rightarrow\pi^{+}\nu\bar{\nu})$ value chosen when extracting $\mathcal{B}(K^+ \to \pi^+ X)$ limits is checked using alternative values in the range  $(7.2-9.0)\times10^{-11}$, 
which covers the 
uncertainties quoted in~\cite{Buras:2022wpw, DAmbrosio:2022kvb}. 
The change in the $\mathcal{B}(K^+ \to \pi^+ X)$ limits is found to be negligible.

The search for peaks in the $m_{\rm miss}^{2}$ distribution uses the $K^{+}\rightarrow\pi^{+}\nu\bar{\nu}$ candidate events, which pass comprehensive veto criteria rejecting any pair of final state particles in addition to the $\pi^{+}$.
Therefore,  
if the $X$ particle decays to visible SM particles within the detector,
the event is rejected from this search
with inefficiency below $10^{-7}$~\cite{Pnn2016paper,Pnn2017paper,PnnRun1Paper,NA62Pnn2122}.  
As a consequence, the highest sensitivity is obtained for $X$ particles with proper lifetime $\tau_X > 1\,\text{ns}$, or scenarios where $X$ decays invisibly.
Model-independent constraints for $\mathcal{B}(K^{+}\rightarrow\pi^{+}X)$ are also established for decays of $X$ into visible particles and with small $\tau_{X}$ values, as shown in figure~\ref{fig:KpiX_1622_limits}-b.
Simulations of $K^{+}\rightarrow\pi^{+}X, \, X\rightarrow\ell^{+}\ell^{-}$ prompt decay chains are used to evaluate the selection acceptance, and therefore the SES according to equation~\ref{eqn:Acc_SES}, for $X$ particles of mass $m_{X}$ and lifetime $\tau_{X}$. Equation~\ref{eqn:BrULs} is used to establish an upper limit for the branching ratio corresponding to each combination of $m_{X}$ and $\tau_{X}$. 
For $\tau_{X}>\mathcal{O}(10\,\text{ns})$ the $X$ particles become sufficiently long-lived to escape the detector, making them indistinguishable from invisibly decaying $X$.  
For $\tau_{X}<\mathcal{O}(100\,\text{ps})$ the $X$ particles decay in the detector so that this search has negligible sensitivity.

\subsection{Search for \boldmath{$\pi^{0}$} decays to invisible final states}
\label{sec:pi0inv}
\begin{figure}[tb]
\centering
\includegraphics[width=0.51\linewidth]{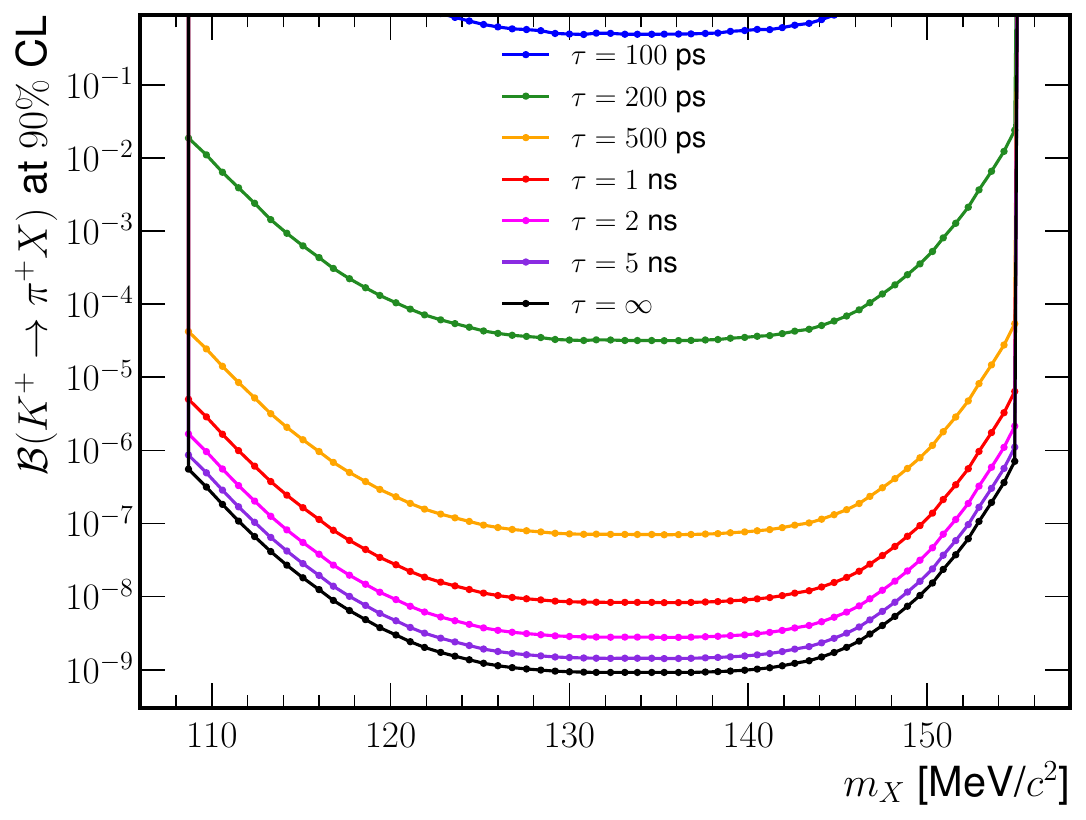}
\caption{
Upper limits at $90\,\%\,\text{CL}$ for $\mathcal{B}(K^+ \to \pi^+ X)$ 
as a function of
$m_X$ for several $\tau_X$ hypotheses, from searches for $\pi^{0}$ decays to invisible final states~\cite{NA62pi0inv}.
}
\label{fig:BR_pi0inv}
\end{figure}

A search for $\pi^{0}$ decays to invisible final states was performed on 2017 data~\cite{NA62pi0inv}, selecting a sample of $K^{+}\rightarrow\pi^{+}\pi^0$ decays using criteria similar to those used for the $K^{+}\rightarrow\pi^{+}\nu\bar{\nu}$ study. 
This search is interpreted to provide model-independent upper limits for 
$\mathcal{B}(K^{+}\rightarrow\pi^{+}X)$, 
where $m_{X}$ is close to the $\pi^{0}$ mass in the range $110$--$155\,\text{MeV}/c^{2}$. 
Limits are placed for a range of $\tau_{X}$ values, as shown in figure~\ref{fig:BR_pi0inv}.

\section{Searches for \boldmath{$K^{+}\rightarrow\pi^{+}X$} with prompt $X$ decays to visible final states}
\label{sec:NA62Measurements_vis}

\subsection{Interpretation of \boldmath{$K^{+}\rightarrow\pi^{+}\mu^{+}\mu^{-}$} results}
Based on 2017--2018 data, NA62 conducted a study of the $K^{+}\rightarrow\pi^{+}\mu^{+}\mu^{-}$ decay~\cite{NA62Kpimumu}.
In scenarios allowing $X\rightarrow\mu^{+}\mu^{-}$ decays, a peak search in the di-muon mass ($m_{\mu\mu}$) spectrum allows constraints to be set on 
$\mathcal{B}(K^{+}\rightarrow\pi^{+}X)\times\mathcal{B}(X\rightarrow\mu^{+}\mu^{-})$, presented in this work for the first time.

\begin{figure}[b]
\centering
\subfloat[]{\includegraphics[width=0.49\linewidth]{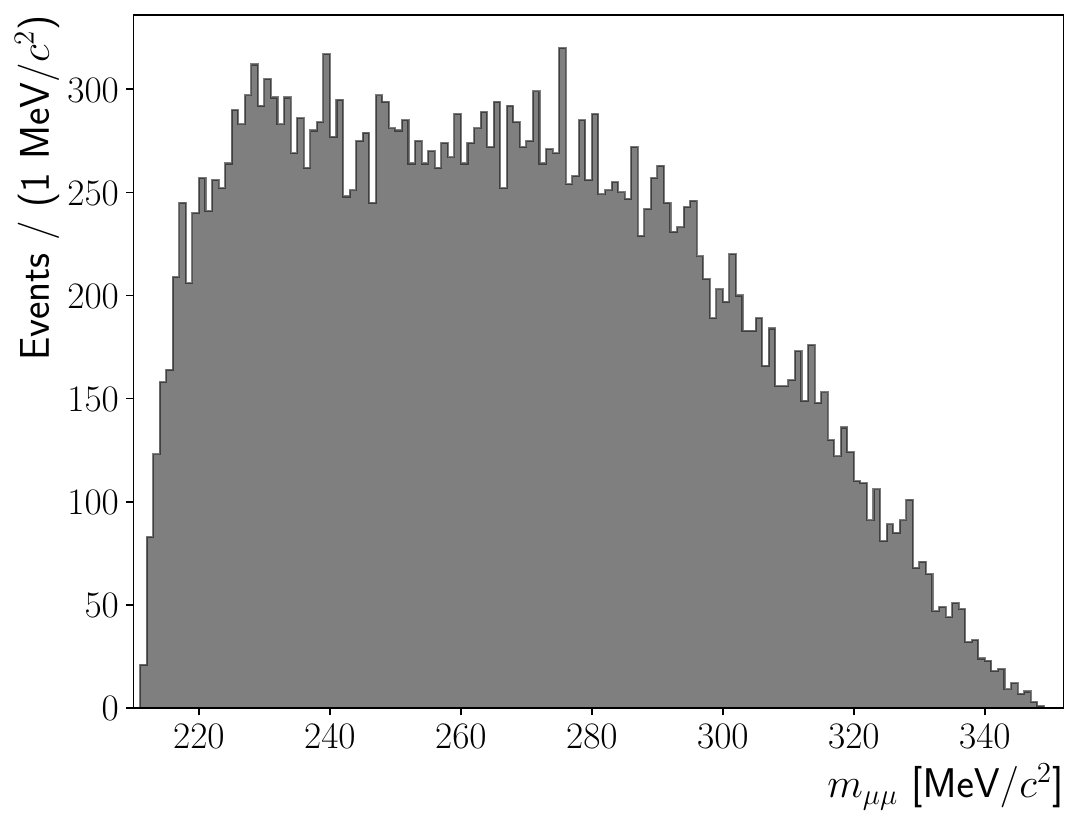}}\
\subfloat[]{\includegraphics[width=0.49\linewidth]{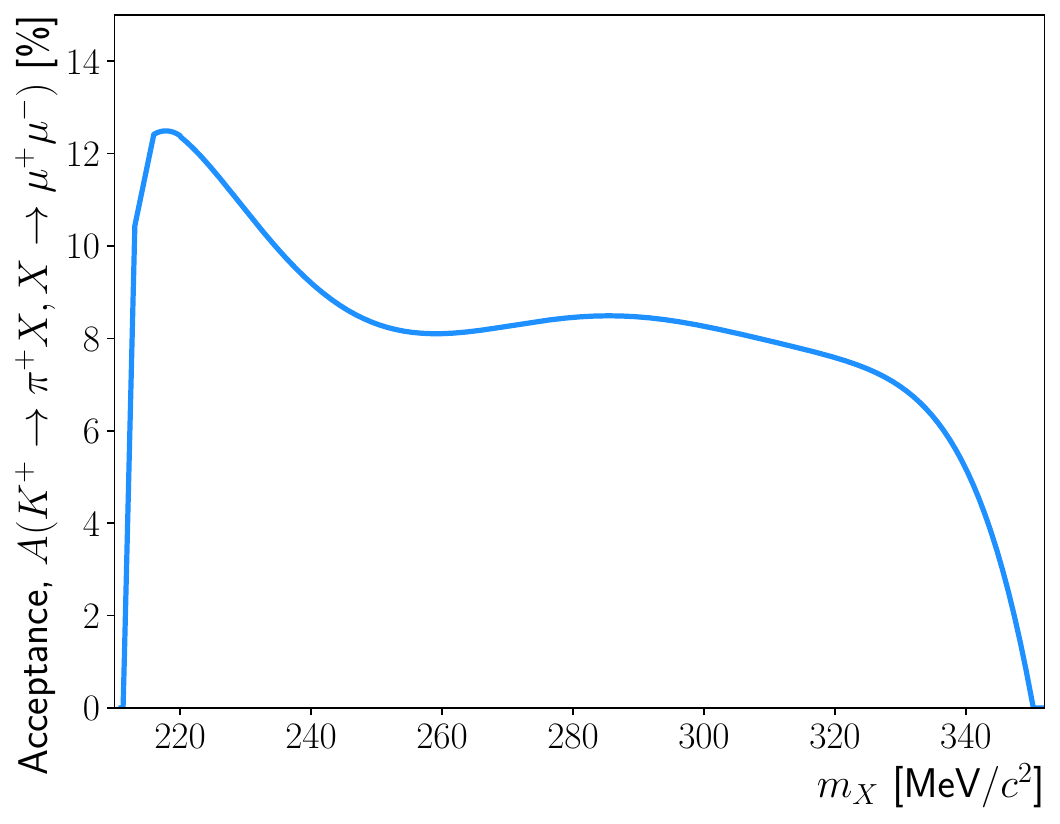}}\
\caption{
(a): Di-muon mass spectrum of selected $K^{+}\rightarrow\pi^{+}\mu^{+}\mu^{-}$ events from 2017--2018 data.
(b): Selection acceptance for the decay chain $K^{+}\rightarrow\pi^{+}X$, $X\rightarrow\mu^{+}\mu^{-}$ from simulations with $\tau_{X}=0$. 
}
\label{fig:KpXmm_Acc}
\end{figure}

\begin{figure}[tb]
\centering
\subfloat[]{\includegraphics[width=0.49\linewidth]{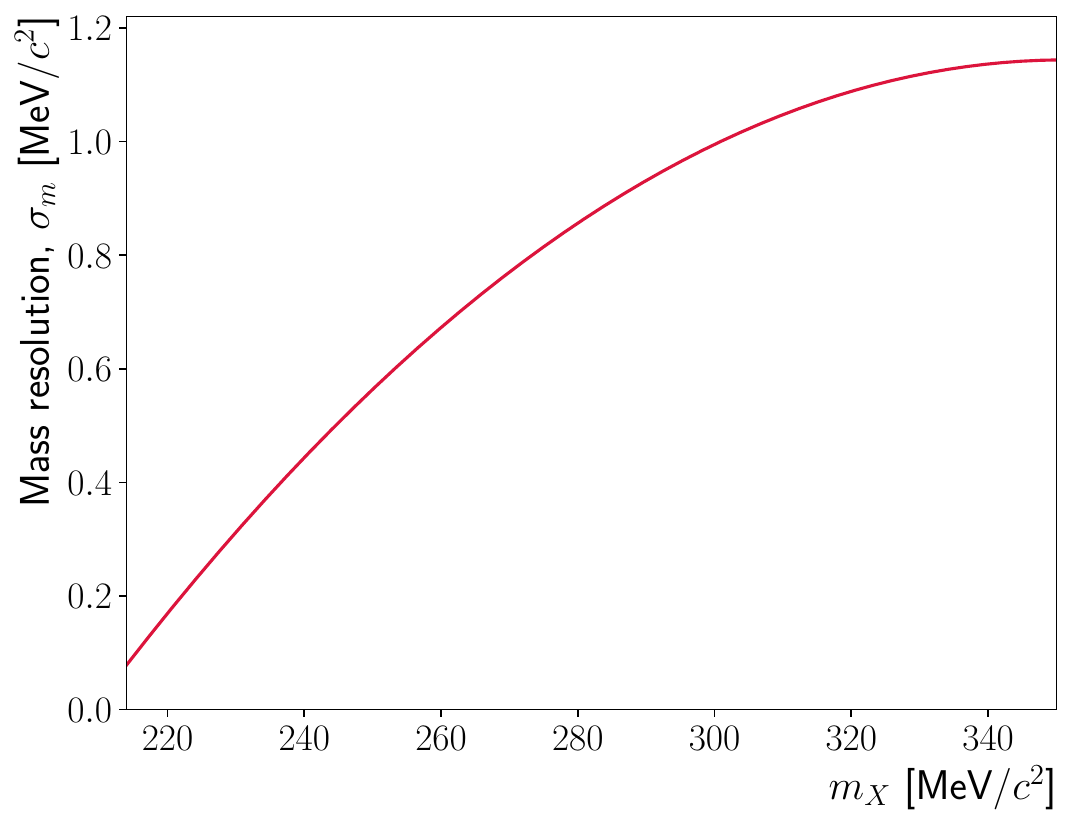}}\
\subfloat[]{\includegraphics[width=0.49\linewidth]{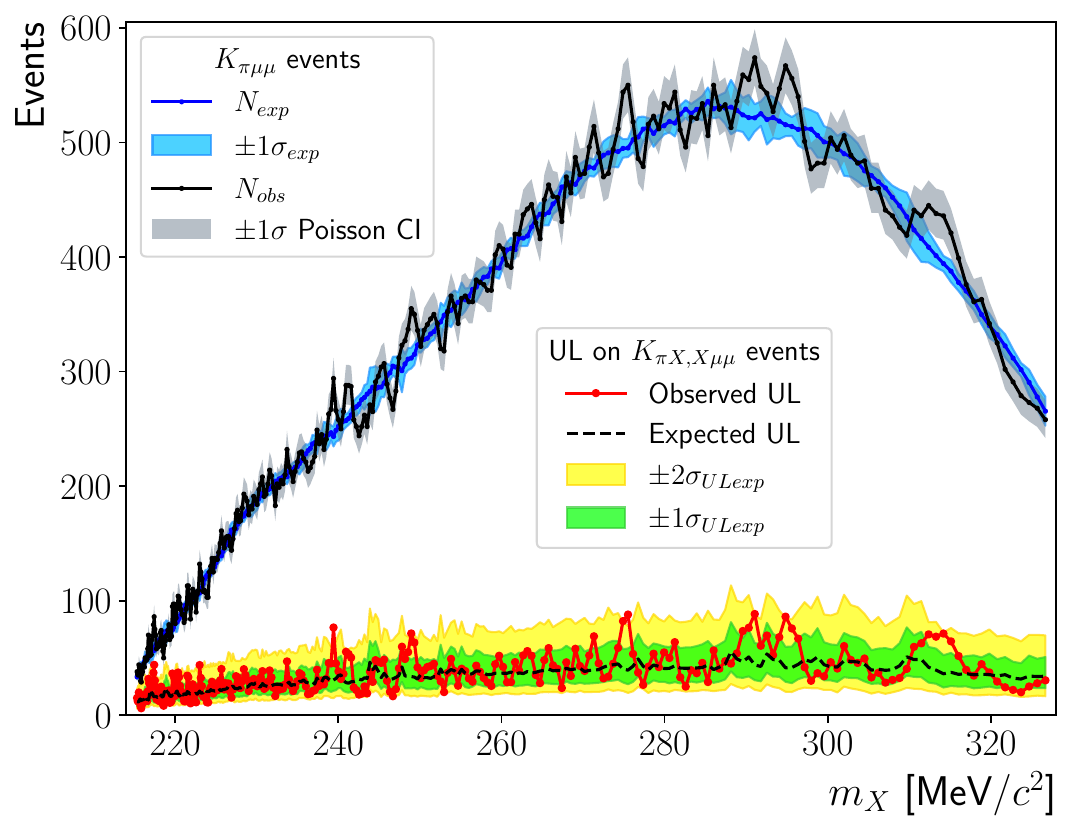}}\
\caption{
(a): Di-muon mass resolution as a function of $m_{X}$.
(b): Numbers of expected (blue) and observed (black) $K^{+}\rightarrow\pi^{+}\mu^{+}\mu^{-}$ events in each search window, together with the $\pm1\sigma$ Poisson confidence interval, CI. The corresponding upper limits at $90\,\%\,\text{CL}$ for the numbers of expected (dashed line) and observed  (red line) $K^{+}\rightarrow\pi^{+}X, \,X\rightarrow\mu^{+}\mu^{-}$ events are also shown.
}
\label{fig:KpXmm_ResAndN}
\end{figure}

In the analysis of~\cite{NA62Kpimumu}, $27679$ candidate $K^{+}\rightarrow\pi^{+}\mu^{+}\mu^{-}$ events are selected, with negligible background. 
The $m_{\mu\mu}$ spectrum observed in the data is shown in figure~\ref{fig:KpXmm_Acc}-a.
The selection acceptance of the decay chain $K^{+}\rightarrow\pi^{+}X$, $X\rightarrow\mu^{+}\mu^{-}$ as a function of $m_{X}$ is evaluated with simulations assuming isotropic $X \to \mu^+ \mu^-$ decays and several $\tau_X$ hypotheses. 
The acceptance obtained for $\tau_{X}=0$ is shown in figure~\ref{fig:KpXmm_Acc}-b as a function of $m_{X}$, 
and is found to decrease  
as a function of $\tau_{X}$ approximately as
\begin{equation}
    A(m_{X},\tau_{X}) = A(m_{X},\tau_{X}=0)\cdot\left( 1 - e^{- \tau_{0}/\tau_{X}} \right) \,\,,
    \label{eqn:KpXmmAcc}
\end{equation}
where $\tau_{0} = 0.07\,\text{ns}$.
A peak search is performed for $298$ di-muon mass hypotheses between $215.3$ and $326.7\,\text{MeV}/c^2$. 
This range is determined by the availability of sidebands with sufficient statistics.
The distance between adjacent mass hypotheses is equal to the di-muon mass resolution $\sigma_{m}$, shown in figure~\ref{fig:KpXmm_ResAndN}-a.
For each hypothesis, a search window of $\pm \, 1\, \sigma_{m}$ is considered.
The background in each window is evaluated from data, using $\pm \, 9 \, \sigma_{m}$ sidebands, and excluding the search window. 
A linear fit to the sideband data provides the expected background in the search window and its statistical uncertainty;
the systematic uncertainty is evaluated using alternative quadratic and cubic polynomial fits. 
The expected and observed numbers of events in each window are shown in figure~\ref{fig:KpXmm_ResAndN}-b.

\begin{figure}[tb]
\centering
\subfloat[]{\includegraphics[width=0.49\linewidth]{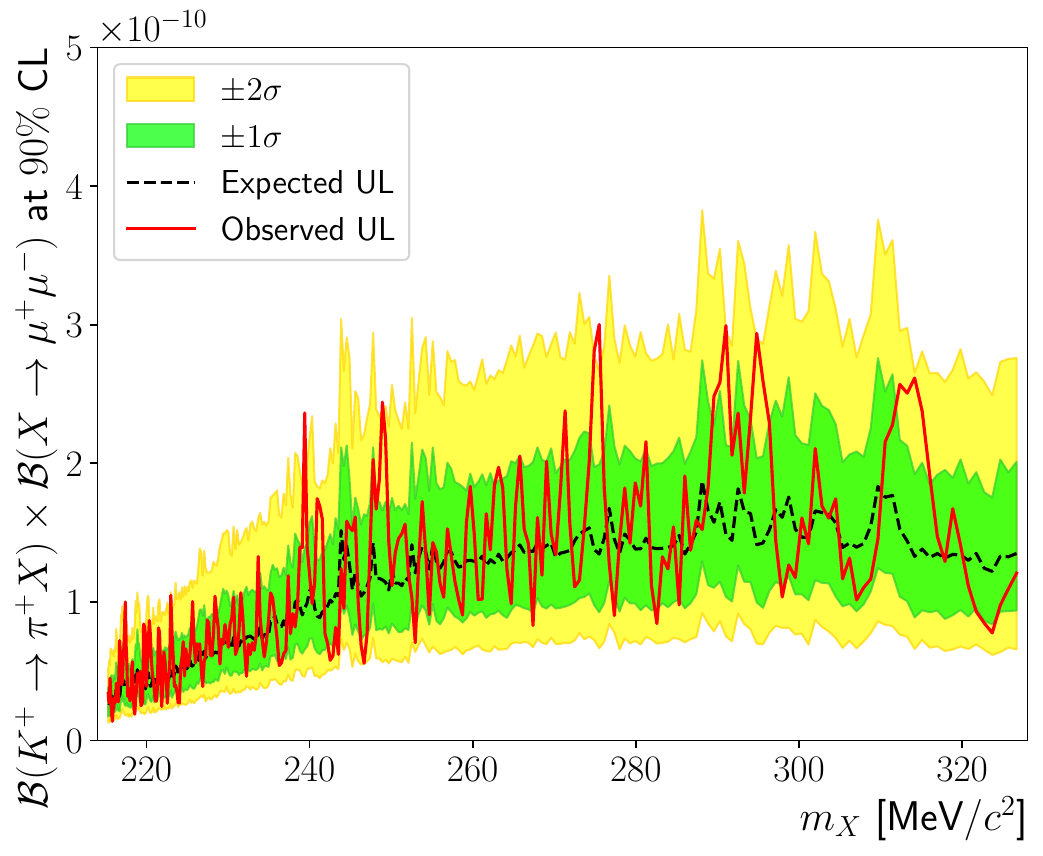}}\
\subfloat[]{\includegraphics[width=0.49\linewidth]{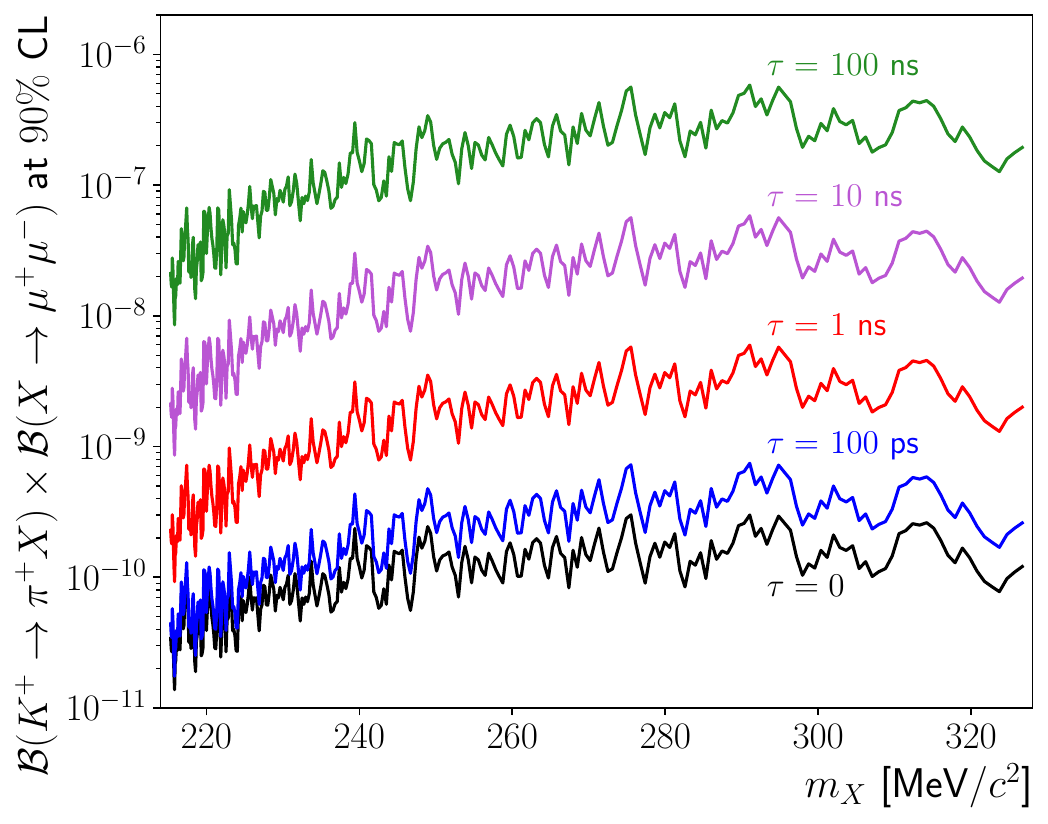}}\
\caption{
(a): Expected and observed upper limits at $90\,\%$ CL for  
$\mathcal{B}(K^{+}\rightarrow\pi^{+}X)\times\mathcal{B}(X\rightarrow\mu^{+}\mu^{-})$ as a function of $m_{X}$ for $\tau_{X}=0$.
(b): Observed upper limits for $\mathcal{B}(K^{+}\rightarrow\pi^{+}X)\times\mathcal{B}(X\rightarrow\mu^{+}\mu^{-})$ at $90\,\%\,\text{CL}$ as a function of $m_{X}$ for several $\tau_X$ values.
}
\label{fig:KpXmm_BRULs}
\end{figure}

No significant excess is observed above the background for any mass hypothesis, with a maximum local significance of $3$~standard deviations.
Upper limits are set on $N_{UL}^{K\pi X, X\mu\mu}$, the number of events of the decay chain $K^{+}\rightarrow\pi^{+}X$, $X\rightarrow\mu^{+}\mu^{-}$ for each mass hypothesis (figure~\ref{fig:KpXmm_ResAndN}-b), using the $\text{CL}_{\text{S}}$ method~\cite{Read:2002hq}. These limits are converted to upper limits for the branching ratio product according to
\begin{equation}
    UL(\mathcal{B}(K^{+}\rightarrow\pi^{+}X)\times\mathcal{B}(X\rightarrow\mu^{+}\mu^{-}))(m_{X},\tau_{X}) = 
    \frac{N_{UL}^{K\pi X, X\mu\mu}}{N_{K} \, A(m_{X},\tau_{X})} \,\,,
\end{equation}
where $A(m_{X},\tau_{X})$ is given by equation~\ref{eqn:KpXmmAcc}
and $N_{K} = (3.48\pm0.09)\times10^{12}$ is the effective number of $K^{+}$ decays in the fiducial decay volume~\cite{NA62Kpimumu}. 
Figure~\ref{fig:KpXmm_BRULs}-a displays the expected and observed limits in the scenario $\tau_{X}=0$, while figure~\ref{fig:KpXmm_BRULs}-b shows observed limits for several $\tau_{X}$ values.

\subsection{Interpretation of \boldmath {$K^{+}\rightarrow\pi^{+}\gamma\gamma$} results}

\begin{figure}[tb]
\centering
\subfloat[]{\includegraphics[width=0.49\linewidth]{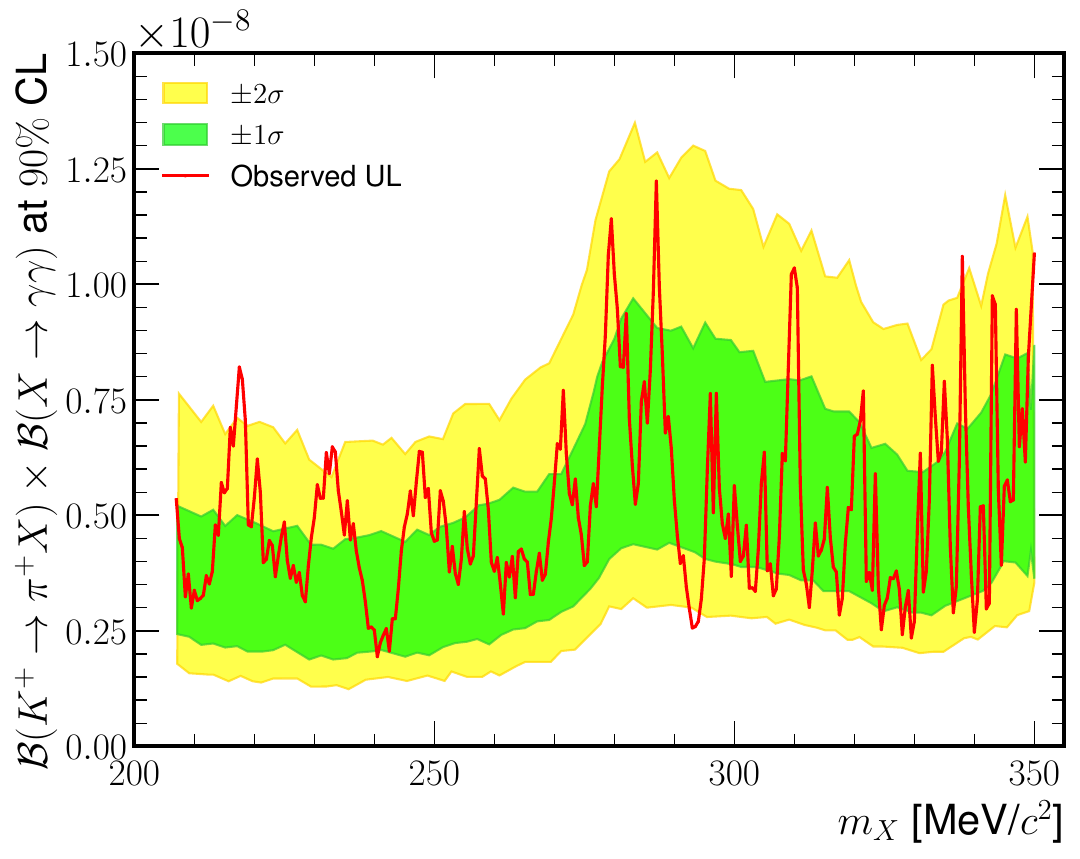}}\
\subfloat[]{\includegraphics[width=0.49\linewidth]{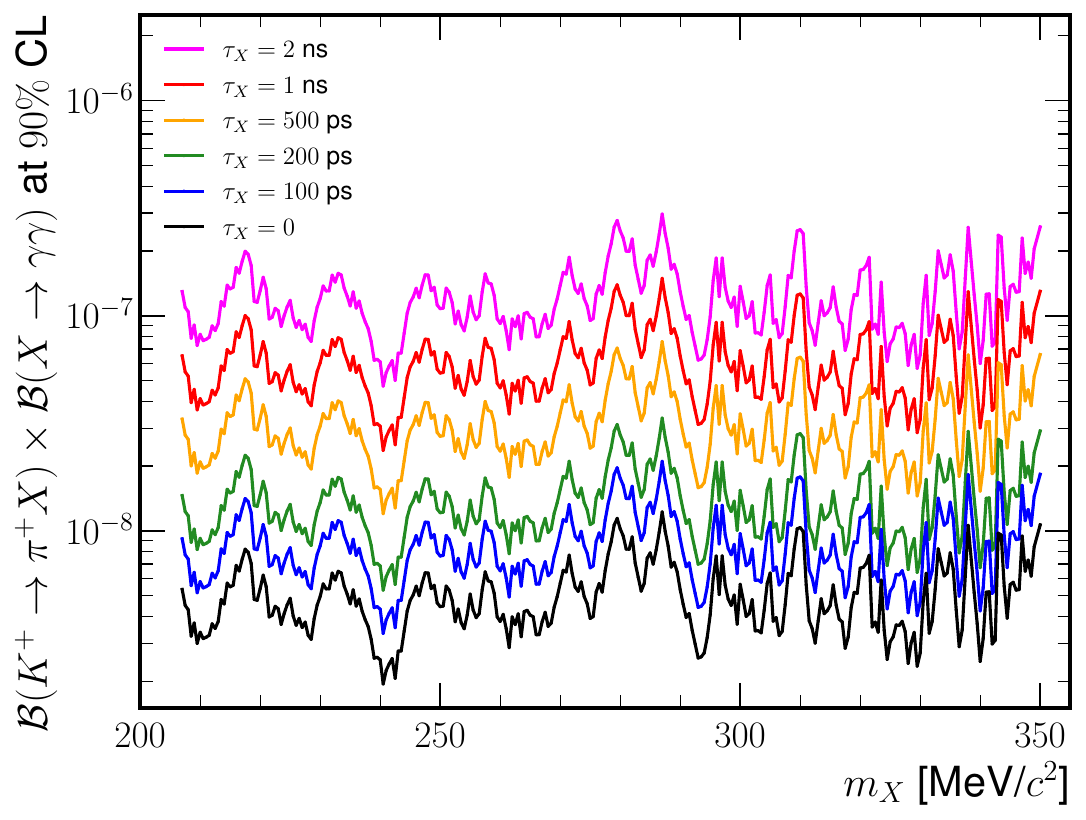}}\
\caption{
(a): Upper limits at $90\,\%\,\text{CL}$ for $\mathcal{B}(K^{+}\rightarrow\pi^{+}X)\times\mathcal{B}(X\rightarrow\gamma\gamma)$ as a function of $m_{X}$ for $\tau_{X}=0$~\cite{NA62:2023olg}.
(b): Upper limits at $90\,\%\,\text{CL}$ for $\mathcal{B}(K^{+}\rightarrow\pi^{+}X)\times\mathcal{B}(X\rightarrow\gamma\gamma)$ for several $\tau_X$ values.
}
\label{fig:BR_Kpigg}
\end{figure}

Using 2017--2018 data, NA62 performed a measurement of the SM $K^{+}\rightarrow\pi^{+}\gamma\gamma$ decay~\cite{NA62:2023olg}.
A peak search in the $m_{\gamma\gamma}$ spectrum obtained limits for 
$\mathcal{B}(K^{+}\rightarrow\pi^{+}X)\times\mathcal{B}(X\rightarrow\gamma\gamma)$,
shown in figure~\ref{fig:BR_Kpigg}-a. 
These limits become weaker for larger $\tau_{X}$, as shown in figure~\ref{fig:BR_Kpigg}-b, 
since the acceptance decreases with a more displaced $X\rightarrow\gamma\gamma$ vertex. 
The sensitivity of this search is limited to $\tau_{X} < 3 \, \text{ns}$.

\section{Constraints on benchmark hidden-sector models}
\label{sec:DarkSectorConstraints}

Using the model-independent limits described in sections~\ref{sec:NA62Measurements_inv} and~\ref{sec:NA62Measurements_vis}, constraints are established on benchmark hidden-sector models. 
Depending on the scenario, the $X$ particle may decay invisibly to dark-sector particles or neutrinos, or visibly to a pair of SM particles.
The lifetime expected for a given mass is determined by the coupling strength to SM fields, and the number of available decay channels. 
Since the model-independent constraints depend on $\tau_{X}$, a certain range of coupling strengths are excluded for a given $m_{X}$.
Constraints on $\mathcal{B}(K^{+}\rightarrow\pi^{+}X)$ from studies of  $K^{+}\rightarrow\pi^{+}\mu^{+}\mu^{-}$ and $K^{+}\rightarrow\pi^{+}\gamma\gamma$ decays also depend on the model-dependent branching ratios of the $X\rightarrow\mu^{+}\mu^{-}$ and $X\rightarrow\gamma\gamma$ decays.

The benchmark hidden-sector models considered in this work are portal models where an additional Lagrangian term, $\mathcal{L_{\rm portal}}$, is introduced to define interactions between hidden-sector and SM fields.
\begin{itemize}
    \item In \textbf{BC1} and \textbf{BC2} a new $U(1)$ symmetry gauge boson $A^{\prime}$, called the dark photon, interacts with the SM fields through kinetic mixing: 
    $\mathcal{L}_{\rm portal} = -(\varepsilon/(2\cos\theta_{W})) F^{\prime}_{\mu\nu}B^{\mu\nu}$,
    where $F^{\prime}_{\mu\nu}$ and $B_{\mu\nu}$ are the field strength tensors of the dark photon and SM hypercharge gauge boson, respectively; $\theta_{W}$ is the Weinberg angle; and the mixing parameter $\varepsilon$ defines the interaction strength.
    \item In \textbf{BC4} a new scalar singlet $S$, called the dark scalar, interacts with the SM Higgs doublet $H$: $\mathcal{L}_{\rm portal} = - \mu S H^{\dagger}H$, where $\mu$ is a coupling constant. 
    Below the electroweak symmetry breaking scale, $S$ mixes with the SM Higgs boson $h$ in proportion to the parameter $\theta \simeq \mu v/(m_{h}^{2}-m_{S}^{2})$, where $v$ is the vacuum expectation value of the Higgs field.
    \item \textbf{BC10} and \textbf{BC11} belong to a class of models where an axion-like particle, $a$, couples to SM fermions, $f$, and gauge bosons, $V$: 
    $\mathcal{L}_{{\rm portal},f} = (C_{ff}/(2\Lambda)) \partial_{\mu}a\bar{f}\gamma^{\mu}\gamma^{5}f$
    and 
    $\mathcal{L}_{{\rm portal,}V} = g^{2} (C_{VV}/\Lambda) a V_{\mu\nu}\tilde{V}^{\mu\nu}$, where $g$ is the corresponding SM gauge boson coupling constant, $\Lambda$ is the new physics energy scale (assumed to be $1\,\text{TeV}$) and $C_{ff}$ and $C_{VV}$ are the coupling constants.
\end{itemize}
In all the scenarios considered, the branching ratio of the $K^+ \to \pi^+ X$ decay is given by 
\begin{equation}
   \mathcal{B}(K^+ \rightarrow \pi^+ X) = \frac{p_X}{8\pi \Gamma_K m_K^2} |\mathcal{M}|^2 \ ,
   \label{eqn:BpiXgeneral}
\end{equation}
where 
$\Gamma_{K}=5.32\times 10^{-14}\, \text{MeV}$ is the $K^{+}$ decay width, 
$p_X$ is the momentum of $X$ in the kaon rest frame, 
and $m_{K}$ is the $K^{+}$ mass~\cite{PDG}.
The matrix element $\mathcal{M}$ depends on the hidden-sector scenario and is proportional to the coupling strength.

\subsection{Massive dark vector} 
\label{sec:BC2}
In QED-like theories~\cite{Okun:1982xi}, a new vector particle (the dark photon $A'$) is introduced, which couples to the electromagnetic current.
This particle with mass $m_{A'}$ mediates interactions both with the SM fields, through a kinetic mixing coupling $\varepsilon\ll1$, and with the hidden-sector fields, through a coupling constant $\alpha_D$ of $\mathcal{O}(1)$.

In the minimal vector portal implementation, the BC1 model~\cite{Beacham:2019nyx}, $A'$ is the only light BSM particle, and is forced to decay exclusively to SM particles.
In this case $K^{+}\rightarrow\pi^{+}X_{\rm inv}$ searches have no sensitivity because $A^{\prime}$ is short-lived and $K^{+}\rightarrow\pi^{+}\mu^{+}\mu^{-}$ searches, which provide limits of $\varepsilon$ of $\mathcal{O}(10^{-3})$, are not competitive~\cite{NA482:2015wmo,LHCb:2019vmc}.
In contrast, in the BC2 model a dark fermion $\chi$ with mass $m_{\chi}<m_{A^{\prime}}/2$ is introduced, and therefore $A'$ decays predominantly to invisible particles, $A^{\prime}\rightarrow\chi\bar{\chi}$. 
The dark fermion is considered to be stable or extremely long-lived and can contribute to the dark matter abundance.
Results from $K^{+}\rightarrow\pi^{+}A^{\prime}$ searches are independent of $\alpha_{D}$ and $m_{\chi}$, and are therefore presented in the ($m_{A’}$, $\varepsilon$) plane.

\begin{figure}[h]
\centering
\subfloat[]{\includegraphics[width=0.49\linewidth]{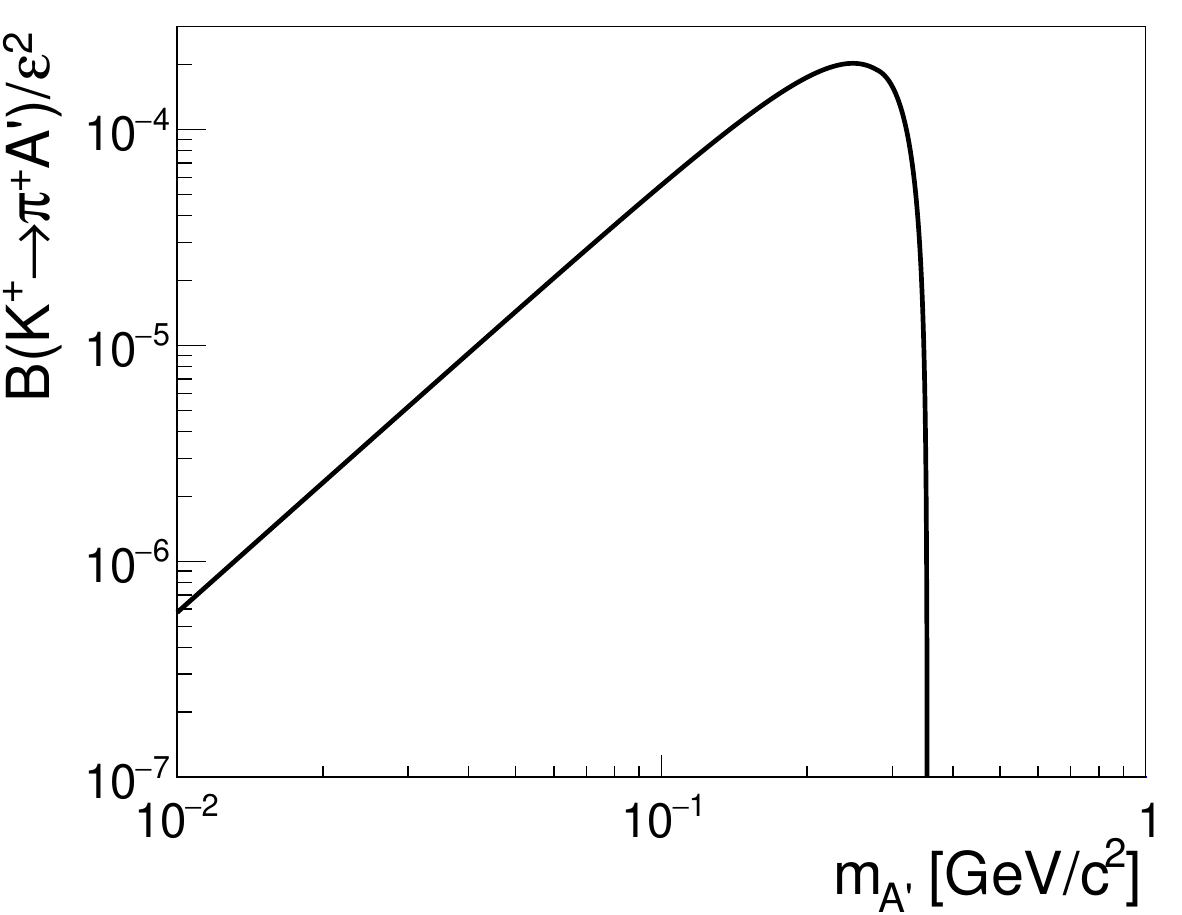}}\
\subfloat[]{\includegraphics[width=0.49\linewidth]{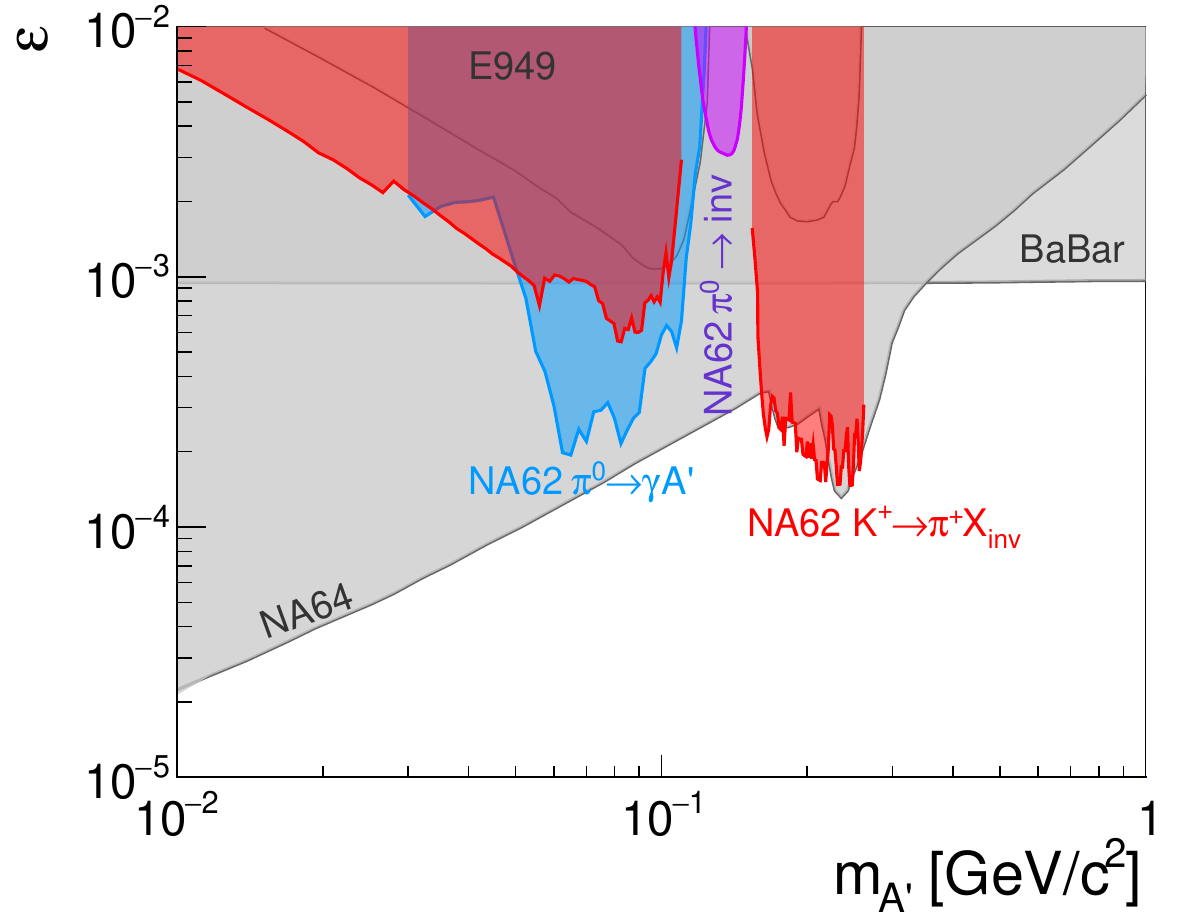}}\
\vspace{-5pt}
\caption{
(a): Branching ratio of the $K^+ \to \pi^+ A'$ decay divided by the kinetic mixing coupling squared, $\varepsilon^{2}$, as a function of $m_{A'}$~\cite{Davoudiasl:2014kua}, according to equation~\ref{eqn:Kdphot}.
(b): Excluded regions, at $90\,\%$ CL,  of the parameter space $(m_{A}^{\prime},\varepsilon)$  
for a dark photon $A^{\prime}$, decaying invisibly, 
in the BC2 model~\cite{Belle:2009zue}.
Excluded regions from NA62 searches for $K^{+}\rightarrow\pi^{+}X_{\text{inv}}$ (figure~\ref{fig:KpiX_1622_limits}),  
$\pi^{0}\rightarrow\text{inv}$~\cite{NA62pi0inv} and $\pi^{0}\rightarrow\gamma A^{\prime}$~\cite{NA62:2019meo} are shown in red, purple and blue, respectively.
Other bounds, shown in grey, are from the experiments 
E949~\cite{BNL-E949:2009dza,Davoudiasl:2014kua},
NA64~\cite{NA64:2023wbi,NA64:2025rib},
and BaBar~\cite{BaBar:2017tiz}.
}
\label{fig:BC2_Inv}
\end{figure}

The radiative decay $K^+ \to \pi^+ \gamma$ violates angular momentum conservation because the photon is massless.
However, the $K^+ \to \pi^+ A'$ decay is allowed if $m_{A'}>0$, although the decay rate is suppressed for small values of $m_{A^{\prime}}$.
In this scenario, the branching ratio of the $K^+ \to \pi^+ A'$ decay is given by equation~\ref{eqn:BpiXgeneral} with a matrix element~\cite{Davoudiasl:2014kua,Pospelov:2008zw}
\begin{equation}
    |\mathcal{M}| = \frac{e\varepsilon |W(z)| m_{A^\prime}}{16 \pi^2 m_K^2 } \sqrt{\lambda(m_K^2, m_\pi^2, m_{A^{\prime}}^2)} \,\,,
\label{eqn:Kdphot}
\end{equation}
where $e$ is the elementary charge, $\lambda$ is the triangle function, $m_{\pi}$ is the charged pion mass, and
$|W(z)|$, introduced in~\cite{DAmbrosio:1998gur}, is evaluated at $z = (m_{A^{\prime}}/m_{K})^{2}$ using the model derived from the NA62 study of the $K^{+}\rightarrow\pi^{+}\mu^{+}\mu^{-}$ decay~\cite{NA62Kpimumu}. 
Figure~\ref{fig:BC2_Inv}-a displays values of $\mathcal{B}(K^+ \to \pi^+ A')/\varepsilon^{2}$ as a function of $m_{A^{\prime}}$.

Using upper limits for $\mathcal{B}(K^{+}\rightarrow\pi^{+}X)$ 
described in section~\ref{sec:NA62Measurements_inv} and equations~\ref{eqn:BpiXgeneral} and~\ref{eqn:Kdphot}, new regions are excluded at $90\,\%$ CL in the $(m_{A^{\prime}},\varepsilon)$ plane. 
For $m_{X}<m_{\pi^{0}}$ and $m_{X} \approx m_{\pi^{0}}$ limits have also been established in this scenario by the NA62 searches for $\pi^{0}\rightarrow\gamma A^{\prime}$~\cite{NA62:2019meo} and $\pi^{0}\rightarrow\text{invisible}$ (section~\ref{sec:pi0inv}), respectively.
The results are displayed in figure~\ref{fig:BC2_Inv}-b, together with limits from other experiments,  
and provide the strongest limits of the $A'$ invisible decays in the $m_{A^{\prime}}$ range $160$--$230$~MeV/$c^2$.

\subsection{Dark scalar mixing with the Higgs boson} 

\begin{figure}[tb]
\centering
\subfloat[]{\includegraphics[width=0.49\linewidth]{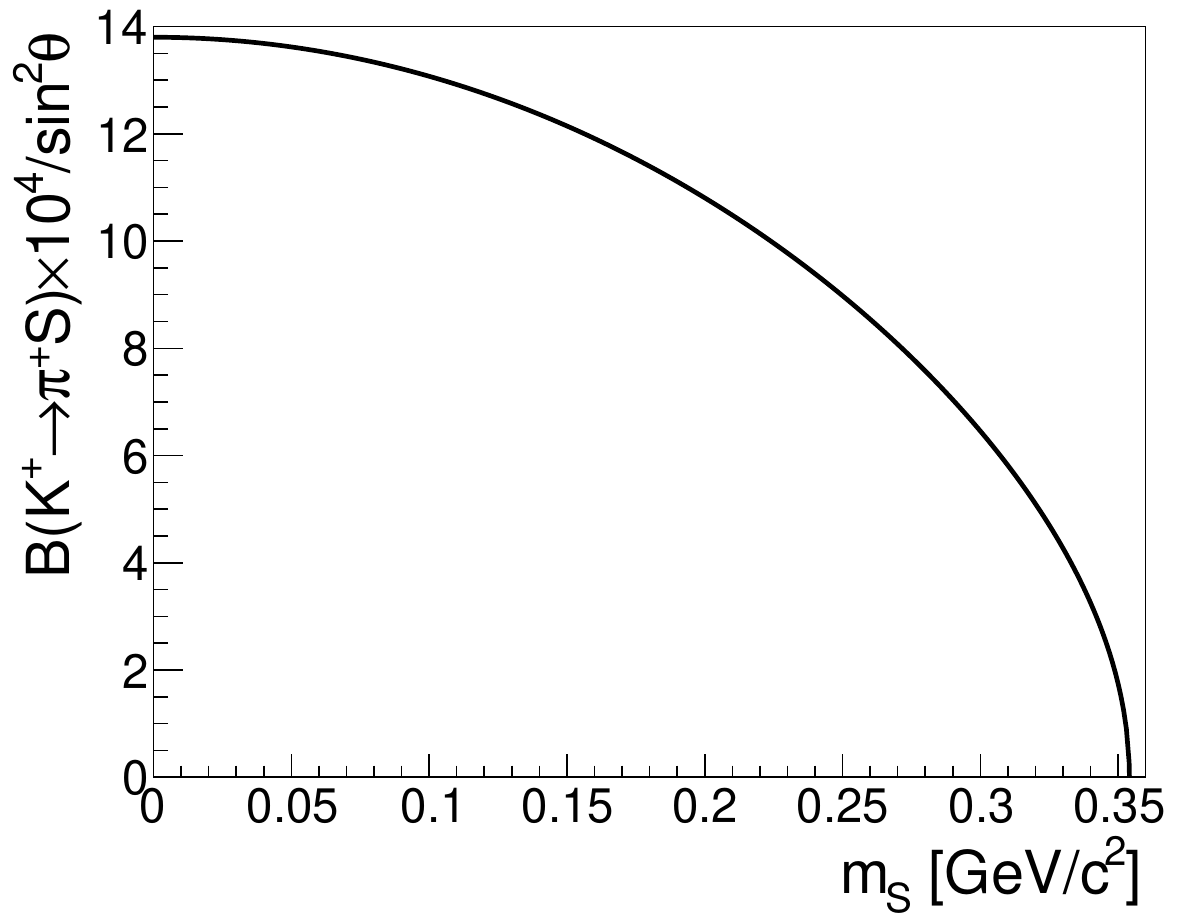}}\
\subfloat[]{\includegraphics[width=0.49\linewidth]{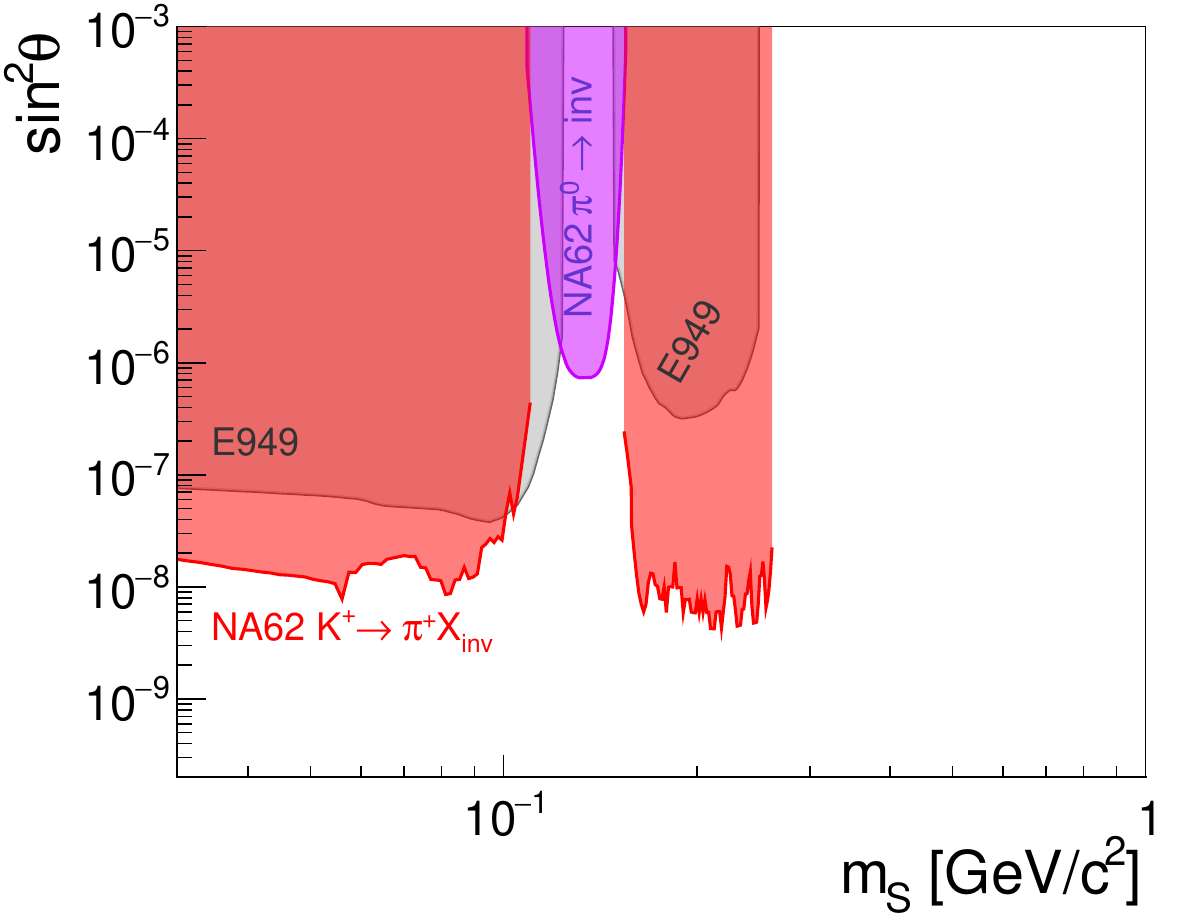}}\
\subfloat[]{\includegraphics[width=0.49\linewidth]{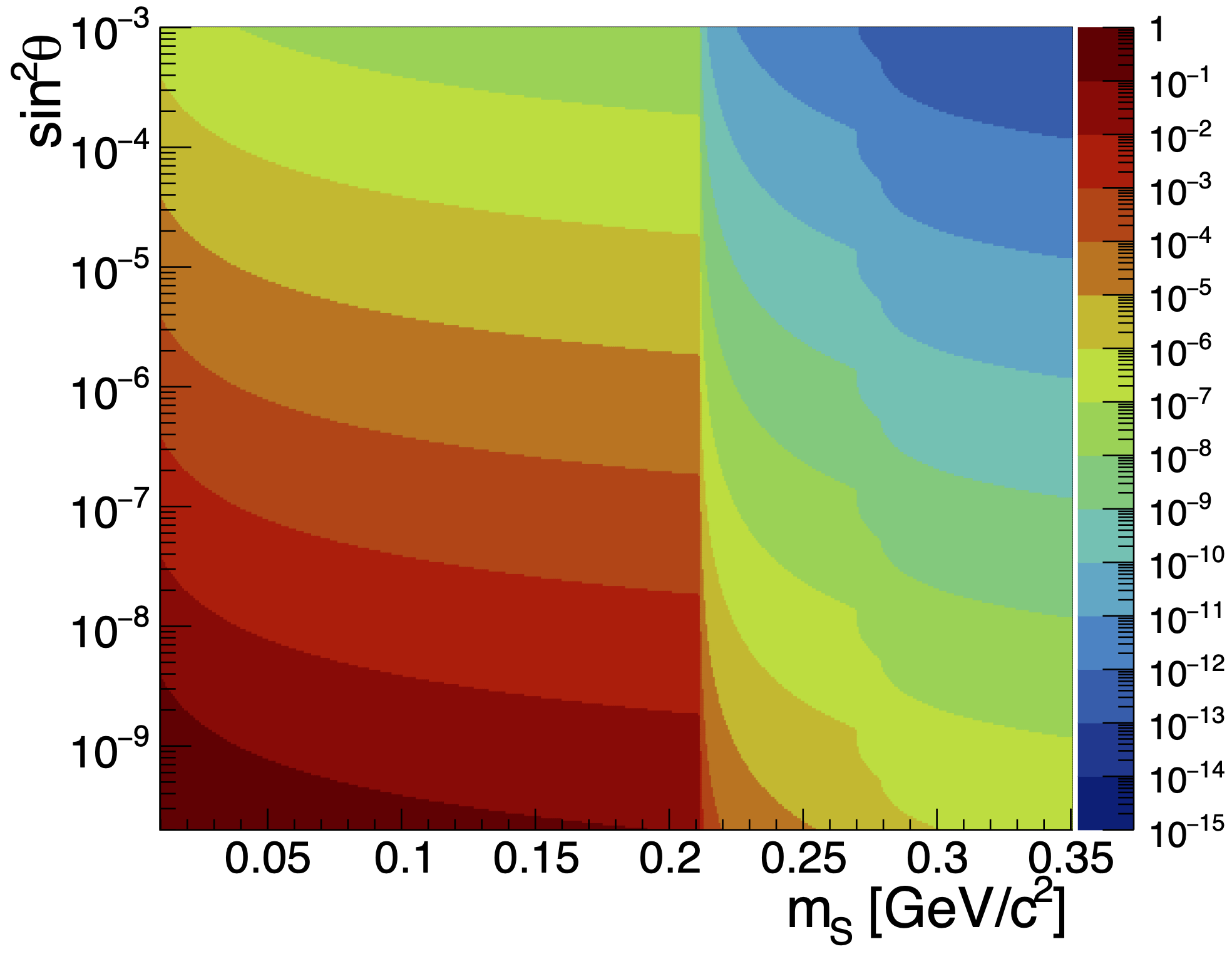}}\
\subfloat[]{\includegraphics[width=0.49\linewidth]{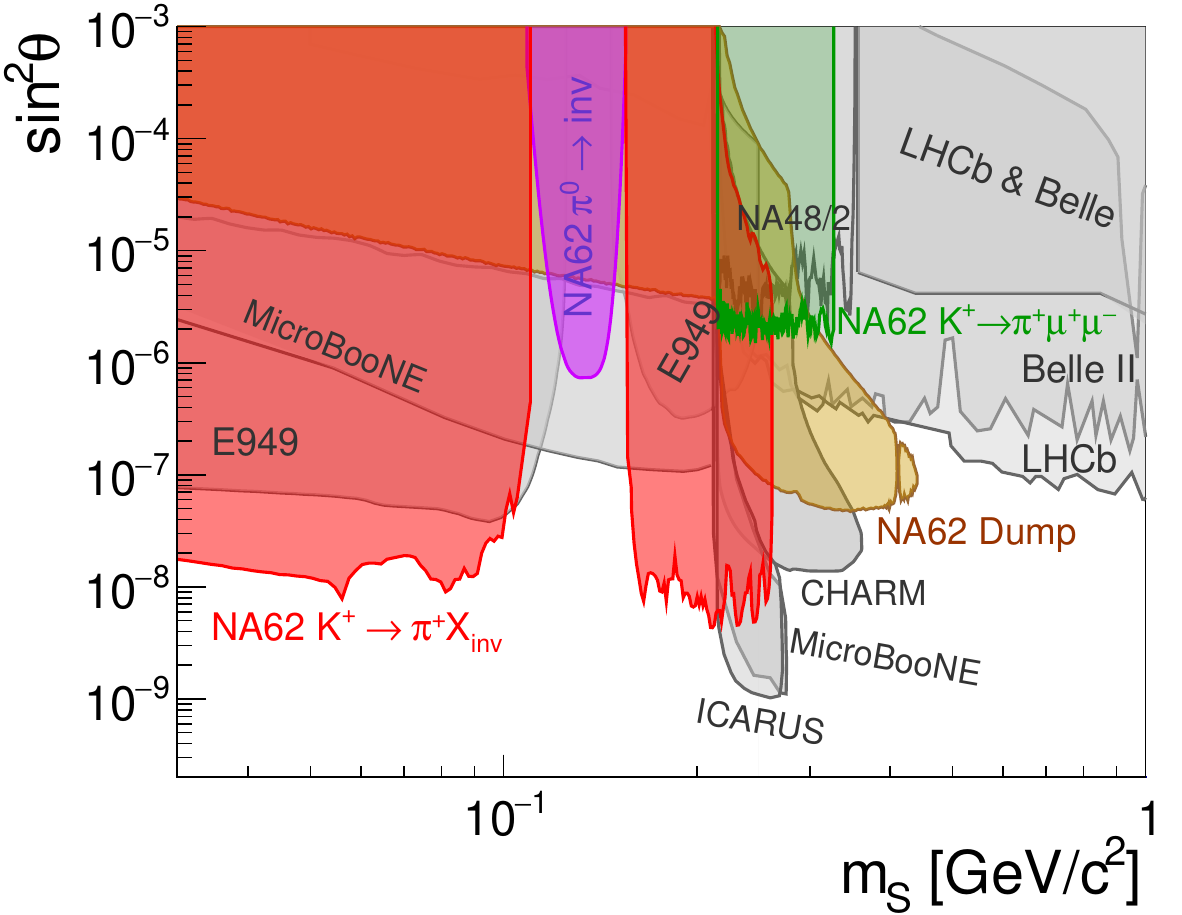}}\
\vspace{-5pt}
\caption{
(a): Branching ratio of the $K^{+}\rightarrow\pi^{+}S$ decay divided by $\sin^{2}\theta$, as a function of $m_S$ (equation~\ref{eqn:BC4_Br}).
(b): Excluded regions, at $90\,\%$ CL, of the parameter space $(m_{S},\sin^{2}\theta)$ for a dark scalar $S$, in the BC4-inv model.
(c): Lifetime (colour scale, $\tau_{S}$ in s) of a dark scalar $S$, in the BC4 model~\cite{Beacham:2019nyx}.
(d): Excluded regions, at $90\,\%$ CL, of the parameter space $(m_{S},\sin^{2}\theta)$ for a dark scalar $S$, 
in the BC4 model~\cite{Beacham:2019nyx}.
Excluded regions from NA62 searches 
for $K^{+}\rightarrow\pi^{+}X_{\text{inv}}$ (figure~\ref{fig:KpiX_1622_limits}), 
$\pi^{0}\rightarrow\text{inv}$~\cite{NA62pi0inv}, $K^{+}\rightarrow\pi^{+}S,\,S\rightarrow\mu^{+}\mu^{-}$ (figure~\ref{fig:KpXmm_BRULs}),
and in beam dump mode~\cite{NA62DumpMode} are shown in red, purple, green and brown, respectively.
Other bounds, shown in grey, are derived from the experiments 
E949~\cite{BNL-E949:2009dza}, 
CHARM~\cite{Winkler:2018qyg}, 
NA48/2~\cite{NA482:2016sfh}, 
LHCb~\cite{LHCb:2015nkv,LHCb:2016awg},
Belle~\cite{Belle:2009zue},
Belle II~\cite{Belle-II:2023ueh},
MicroBooNE~\cite{MicroBooNE:2021sov,MicroBooNE:2022ctm}, and
ICARUS~\cite{ICARUS:2024oqb}.
}
\label{fig:ExlusionBC4}
\end{figure}

In the minimal scalar portal model BC4~\cite{Beacham:2019nyx,Winkler:2018qyg}, hidden-sector particles are coupled to the SM Higgs field. 
A massive scalar particle $S$, with mass $m_S$, mixes with the SM Higgs boson, and both its production and decay (to SM particles) are driven by the mixing parameter $\sin^{2}\theta$. The $K^{+}\rightarrow\pi^{+}S$ branching ratio is given by 
equation~\ref{eqn:BpiXgeneral}, with a matrix element~\cite{Leutwyler:1989xj,Clarke:2013aya}
\begin{equation}
    |\mathcal{M}| = \frac{1}{2} \left( \frac{m_{K}^{2} - m_{\pi}^{2} }{m_{s} - m_{d}}\right) \left( \frac{m_{S}}{v}\frac{3\sqrt{2}G_{F}}{16\pi^{2}}  m_{t}^{2} |V_{ts}^{*}V_{td}| \right) \sin\theta
    = \sqrt{(8\pi \Gamma_K m_K) \, C} \, \sin\theta \,\,,
    \label{eqn:BC4_Br}
\end{equation}
where $G_{F}$ is the Fermi constant, $m_{d,s,t}$ are the masses of the down, strange and top quarks, respectively, and $V_{ij}$ are elements of the CKM matrix. For this analysis $C=3\times10^{-3}$ is used~\cite{Winkler:2018qyg}, while similar values are used elsewhere~\cite{Bezrukov:2009yw,Clarke:2013aya,Feng:2017vli}.
The resulting values of $\mathcal{B}(K^{+}\rightarrow\pi^{+}S)/\sin^{2}\theta$ as a function of $m_S$ are shown in figure~\ref{fig:ExlusionBC4}-a.

In the BC4-inv model,
the dark scalar production phenomenology is identical to BC4 (figure~\ref{fig:ExlusionBC4}-a),
however the dark scalar is stable or decays invisibly to a pair of hidden-sector particles. Upper limits for $\mathcal{B}(K^{+}\rightarrow\pi^{+}X)$ (section~\ref{sec:NA62Measurements_inv}) are converted to upper limits for $\sin^{2}\theta$ according to equations~\ref{eqn:BpiXgeneral} and~\ref{eqn:BC4_Br}.
Results are shown in figure~\ref{fig:ExlusionBC4}-b.

In the BC4 model, where the dark scalar decays to SM particles only and has a mass below the di-pion mass threshold, $S$ decays to lepton pairs $\ell^{+}\ell^{-}$ with a proper lifetime $\tau_{S}$, shown in figure~\ref{fig:ExlusionBC4}-c and given by
\begin{equation}
    \tau_{S}(m_{S},\sin^{2}\theta) = \frac{\hbar}{\Gamma_{S}} \,\,, 
\end{equation}
where the decay width is
\begin{equation}
    \Gamma_{S} = 
    \begin{cases}
        \Gamma(S\rightarrow e^{+}e^{-}) & 2m_{e}\leq m_{S} < 2m_{\mu} \\
        \Gamma(S\rightarrow e^{+}e^{-}) + \Gamma(S\rightarrow \mu^{+}\mu^{-}) & \quad\quad\quad\, m_{S} \geq 2m_{\mu} \,\,,
    \end{cases}
\end{equation}
and
\begin{equation}
    \Gamma(S\rightarrow\ell^{+}\ell^{-}) = 
        \frac{m_{\ell}^{2} m_{S}}{8\pi v^{2}} 
        \left(1 - \frac{4m_{\ell}^{2}}{m_{S}^{2}}\right)^{3/2} 
        \sin^{2}\theta
    \,\,.
\end{equation}
For a given $m_{S}$ value the excluded range of $\sin^{2}\theta$ values depends on $\tau_{S}$. 
Excluded regions from NA62 searches for $K^{+}\rightarrow\pi^{+}X_{\text{inv}}$ (figure~\ref{fig:KpiX_1622_limits}), $\pi^{0}\rightarrow\text{inv}$~\cite{NA62pi0inv} and $K^{+}\rightarrow\pi^{+}S,\,S\rightarrow\mu^{+}\mu^{-}$ (figure~\ref{fig:KpXmm_BRULs}) are displayed in figure~\ref{fig:ExlusionBC4}-d.
As discussed in section~\ref{sec:NA62Measurements_vis}, the sensitivity of NA62 searches is limited by lifetime effects. 
For $m_{S}>2m_{\mu}$ the decay width is dominated by the di-muon decay, therefore the lifetime decreases significantly and the excluded region of the parameter space ($m_S, \sin^2\theta$) is correspondingly reduced (figure~\ref{fig:ExlusionBC4}-d).

\subsection{Axion portal with coupling to SM fermions} 

\begin{figure}[tb]
\vspace{-5pt}
\centering
\subfloat[]{\includegraphics[width=0.49\linewidth]{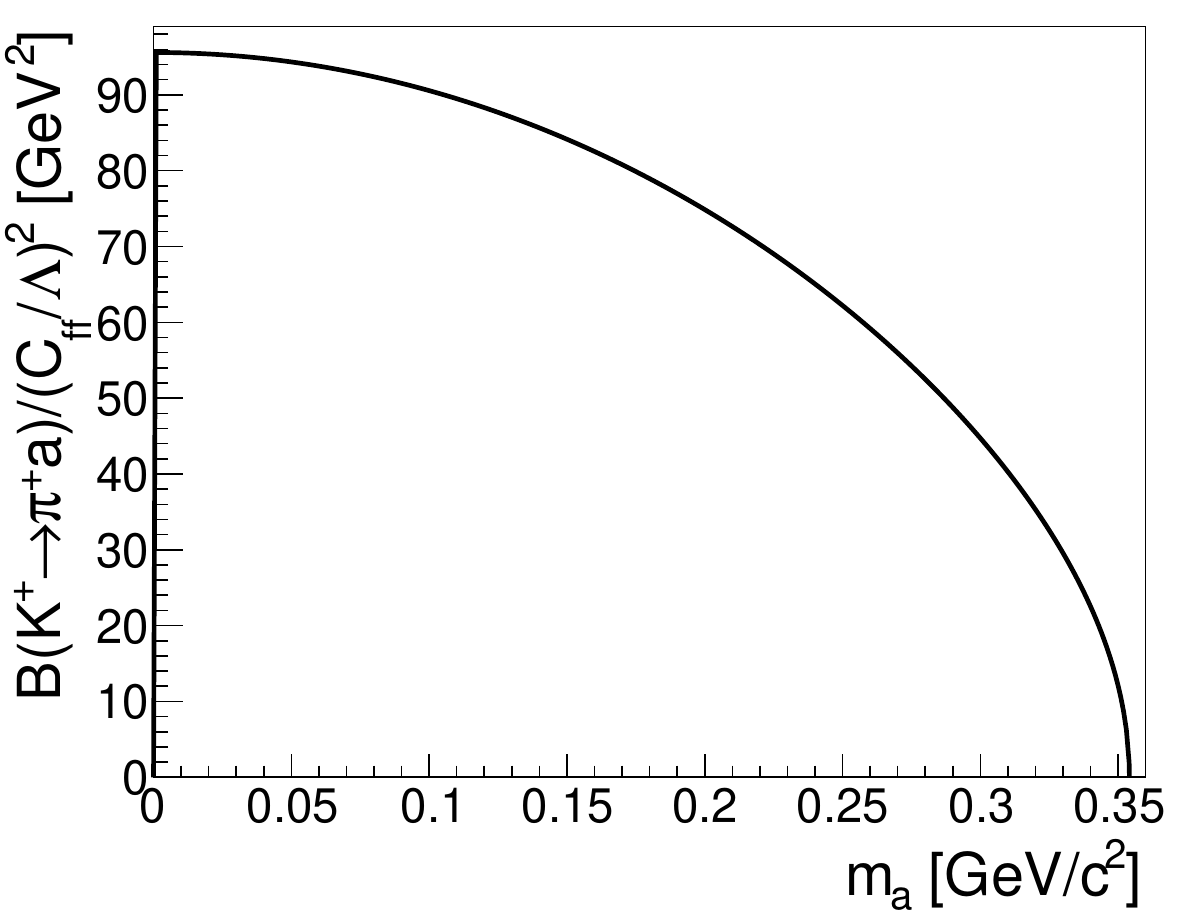}}\
\subfloat[]{\includegraphics[width=0.49\linewidth]{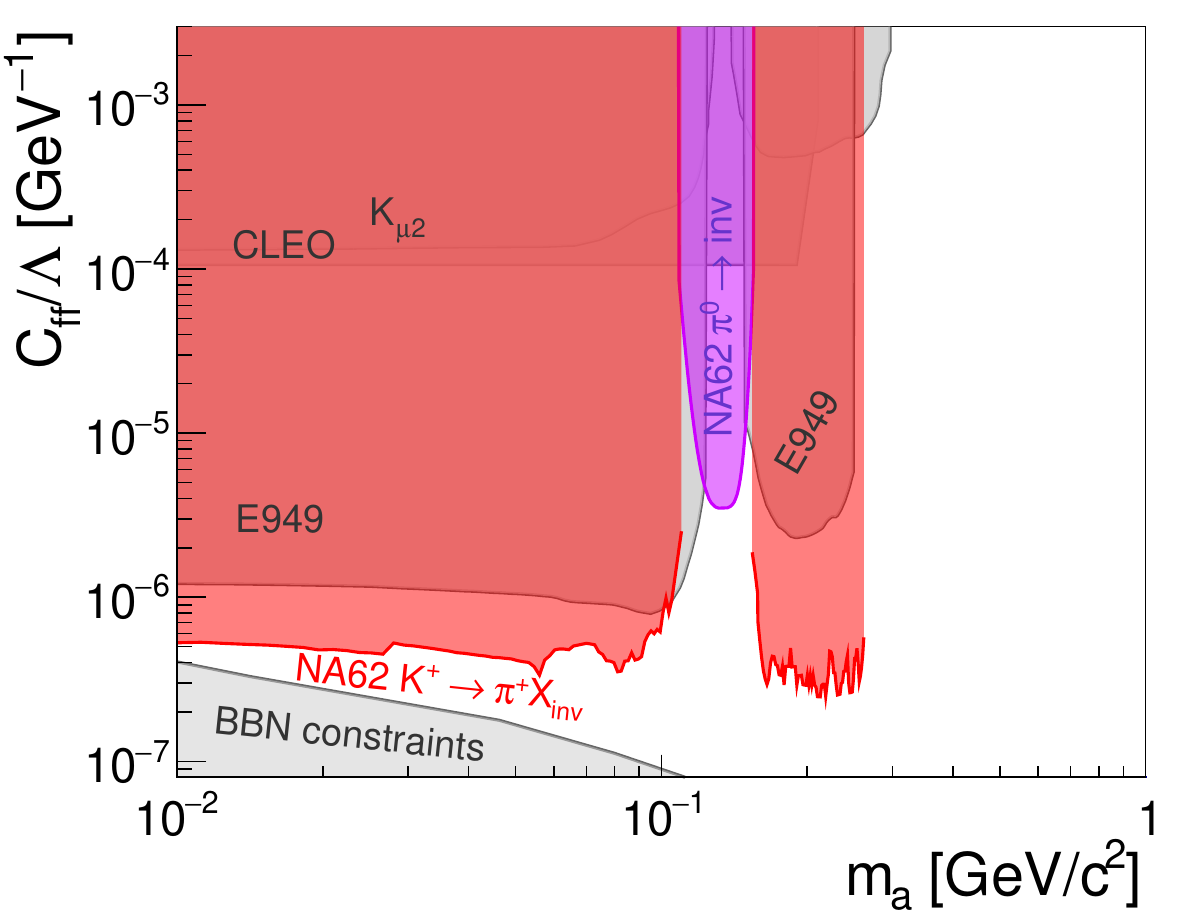}}\
\subfloat[]{\includegraphics[width=0.49\linewidth]{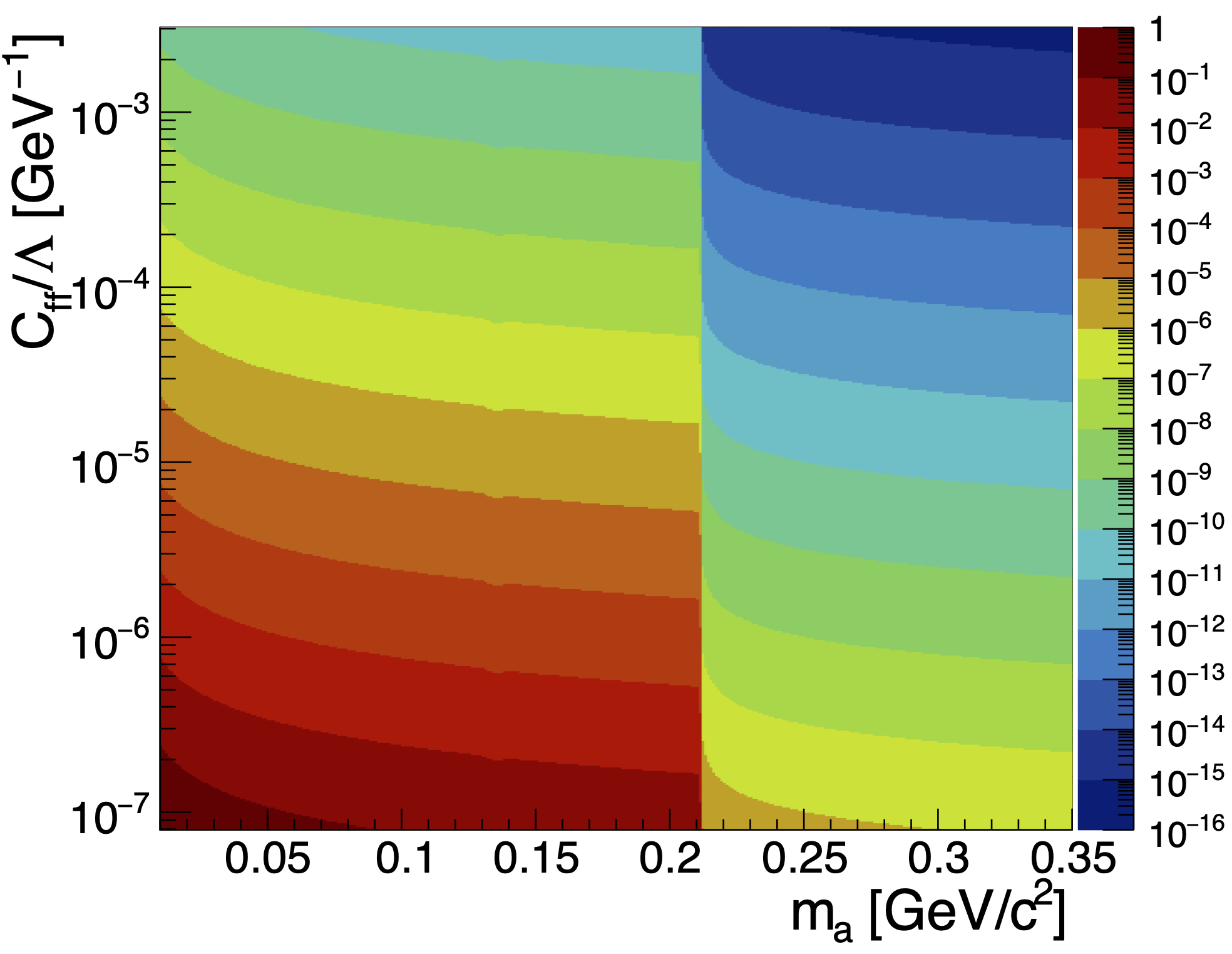}}\
\subfloat[]{\includegraphics[width=0.49\linewidth]{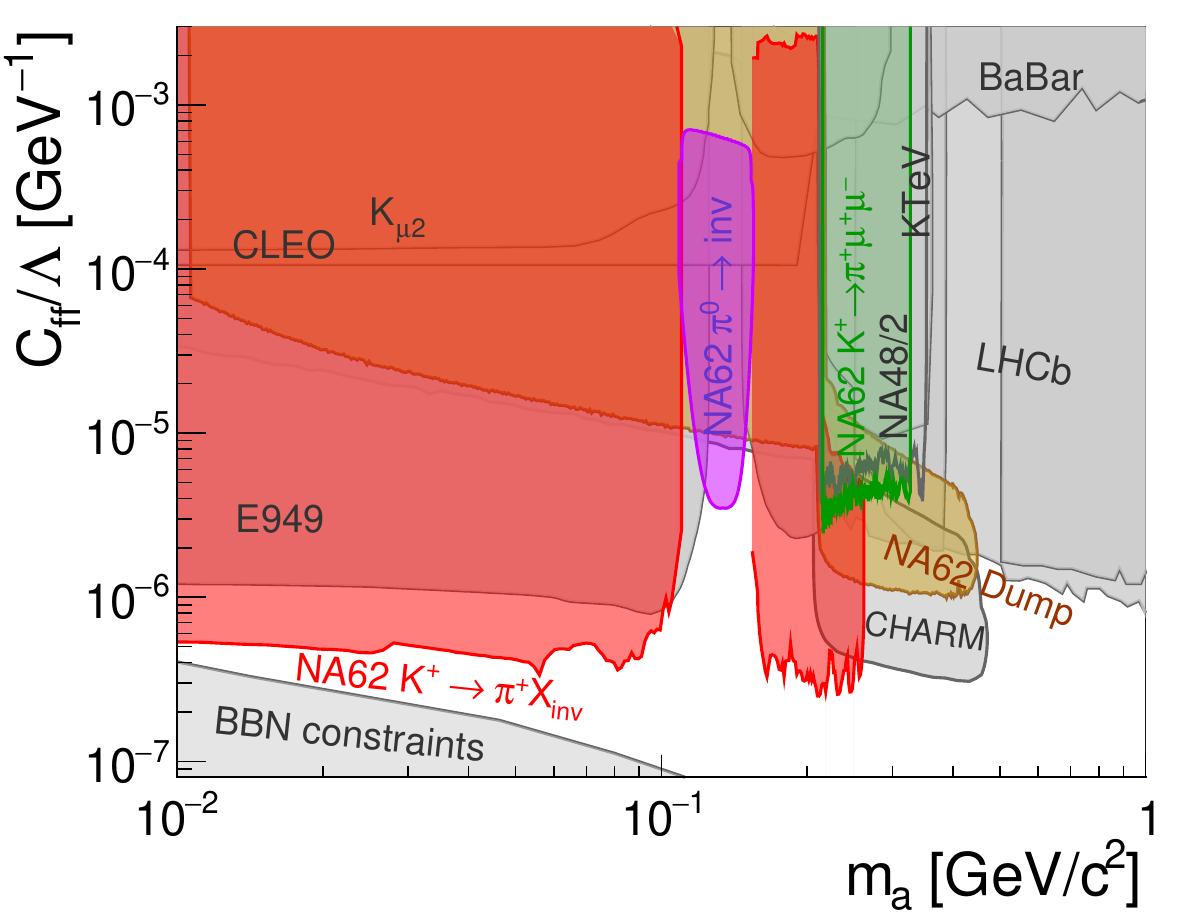}}\
\caption{
(a): Branching ratio of the $K^{+}\rightarrow\pi^{+}a$ decay divided by $(C_{ff}/\Lambda)^{2}$, as a function of $m_{a}$, assuming $\Lambda = 1\,\text{TeV}$~\cite{Jerhot:2022chi}.
(b): Excluded regions, at $90\,\%$ CL, of the parameter space $(m_{a},C_{ff}/\Lambda)$ for an ALP $a$, 
in the BC10-inv model,
evaluated assuming $\Lambda=1\,\text{TeV}$.
(c): Lifetime (colour scale, $\tau_{a}$ in s) of an ALP $a$, in the BC10 model~\cite{Beacham:2019nyx}, as a function of mass and coupling strength~\cite{Jerhot:2022chi}.
(d): Excluded regions, at $90\,\%$ CL, of the parameter space $(m_{a},C_{ff}/\Lambda)$ for an ALP $a$, in the BC10 model~\cite{Beacham:2019nyx}, evaluated assuming $\Lambda=1\,\text{TeV}$.
Excluded regions from NA62 searches
for $K^{+}\rightarrow\pi^{+}X_{\text{inv}}$ (figure~\ref{fig:KpiX_1622_limits}), 
$\pi^{0}\rightarrow\text{inv}$~\cite{NA62pi0inv} and $K^{+}\rightarrow\pi^{+}a,a\rightarrow\mu^{+}\mu^{-}$ (figure~\ref{fig:KpXmm_BRULs}), 
and in beam dump mode~\cite{NA62DumpMode} are shown in red, purple, green and brown, respectively.
Other bounds, shown in grey, are derived from the experiments 
E949~\cite{BNL-E949:2009dza},
$K\mu2$~\cite{Yamazaki:1984vg},
CLEO~\cite{CLEO:2001acz},
CHARM~\cite{CHARM:1985anb},
KTeV~\cite{KTEV:2000ngj},
NA48/2~\cite{NA482:2016sfh},
LHCb~\cite{LHCb:2016awg,LHCb:2015nkv} 
and from Big Bang nucleosynthesis (BBN)~\cite{Beacham:2019nyx}.
}
\label{fig:ExlusionBC10}
\end{figure}

In the axion portal scenario, $a$ is a light pseudoscalar axion-like particle (ALP) which acts as a mediator between the SM and the hidden sector.
A scenario with universal coupling to charged fermions~\cite{Beacham:2019nyx}, parameterized by $C_{ff}$, is considered. 
In this case, unlike the benchmark vector and scalar portal models discussed above, UV completion is required below a certain mass scale, and the renormalization group evolution can introduce additional couplings~\cite{Beacham:2019nyx,Bauer:2020jbp,Bauer:2021mvw}. 
The branching ratio divided by $(C_{ff}/\Lambda)^{2}$ as a function of the ALP mass assuming $\Lambda=1\,\text{TeV}$, based on~\cite{Bauer:2021wjo} and evaluated using the \texttt{ALPINIST} framework~\cite{Jerhot:2022chi}, is displayed in figure~\ref{fig:ExlusionBC10}-a. 

If the ALP is stable or decays invisibly (to a pair of hidden-sector particles, BC10-inv), upper limits for $C_{ff}$ are established according to \texttt{ALPINIST}
from the model-independent $\mathcal{B}(K^{+}\rightarrow\pi^{+}X)$ limits discussed in section~\ref{sec:NA62Measurements_inv}, as shown in figure~\ref{fig:ExlusionBC10}-b.
Alternatively, if the ALP decays to SM particles only (BC10 model) and is light ($m_{a}<2m_{\pi}$), the $a$ lifetime is
\begin{equation}
    \tau_{a}(m_{a},C_{ff})
    = \frac{\hbar}{\Gamma_{\ell\ell} + \Gamma(a\rightarrow\gamma\gamma)}\,\,,
    \label{eqn:LifetimeALP}
\end{equation}
as shown in figure~\ref{fig:ExlusionBC10}-c.
The total $a\rightarrow\ell^{+}\ell^{-}$ decay width is given by
\begin{equation}
    \Gamma_{\ell\ell} =  
    \begin{cases}
        \Gamma(a\rightarrow e^{+}e^{-}) & 2m_{e} \leq m_{a} < 2m_{\mu} \\
        \Gamma(a\rightarrow e^{+}e^{-}) + \Gamma(a\rightarrow \mu^{+}\mu^{-}) & \quad\quad\quad\, m_{a} \geq 2m_{\mu} \,\, ,
    \end{cases}
\end{equation} 
where
\begin{equation}
    \Gamma(a\rightarrow\ell^{+}\ell^{-}) = \left(\frac{C_{ff}}{\Lambda}\right)^{2} \frac{m_{\ell}^{2}m_{a}}{8\pi} \beta_\ell \,\, ,
\end{equation}
$m_{\ell}$ is the lepton mass and $\beta_{\ell}= \sqrt{1-  4m_{\ell}^{2}/m_{a}^{2}}$. 
The width of the decay to a photon pair is given by
\begin{equation}
    \Gamma(a\rightarrow\gamma\gamma) = 
    \left(\frac{C_{ff}}{\Lambda}\right)^{2}\frac{\alpha^{2}m_{a}^{3}}{64\pi^{3}}\left|\sum_{f=e,\mu,u,d,s}N_{C}Q_{f}^{2}C_{ff}F_{A}(\beta_{f}^2)\right|^{2}\,\,,
\end{equation}
where: $f$ labels a fermion; 
$\alpha$ is the fine-structure constant; 
$N_{C}=1 (3)$ for leptons (quarks); 
$Q_{f}$ is the fermion charge; 
$\beta_{u}^2=\beta_{d}^2= ( 1-  4m_{\pi}^{2}/m_{a}^{2})$ and $\beta_{s}^2 =( 1-  4m_{K}^{2}/m_{a}^{2})$; and
\begin{equation}
    F_{A} = \beta_f^2
    \begin{cases}
        \arcsin^{2}\left(\frac{1}{\sqrt{1-\beta_f^2}}\right) & \beta_{f}^2 \leq 0 \\
        -\frac{1}{4}\left[\log\left( \frac{ 1+ \beta_f }{ 1- \beta_f }
        \right) - i\pi \right]^{2} & \beta_{f}^2 > 0   \,\,.
    \end{cases}
\end{equation}
Excluded regions in the $(m_{a},C_{ff}/\Lambda)$ plane from NA62 searches for $K^{+}\rightarrow\pi^{+}X_{\text{inv}}$ (figure~\ref{fig:KpiX_1622_limits}), $\pi^{0}\rightarrow\text{inv}$~\cite{NA62pi0inv} and $K^{+}\rightarrow\pi^{+}S,\,S\rightarrow\mu^{+}\mu^{-}$ (figure~\ref{fig:KpXmm_BRULs}) are displayed in figure~\ref{fig:ExlusionBC10}-d.

\subsection{Axion portal with coupling to SM gluons} 

\begin{figure}[tb]
\vspace{-8pt}
\centering
\subfloat[]{\includegraphics[width=0.51\linewidth]{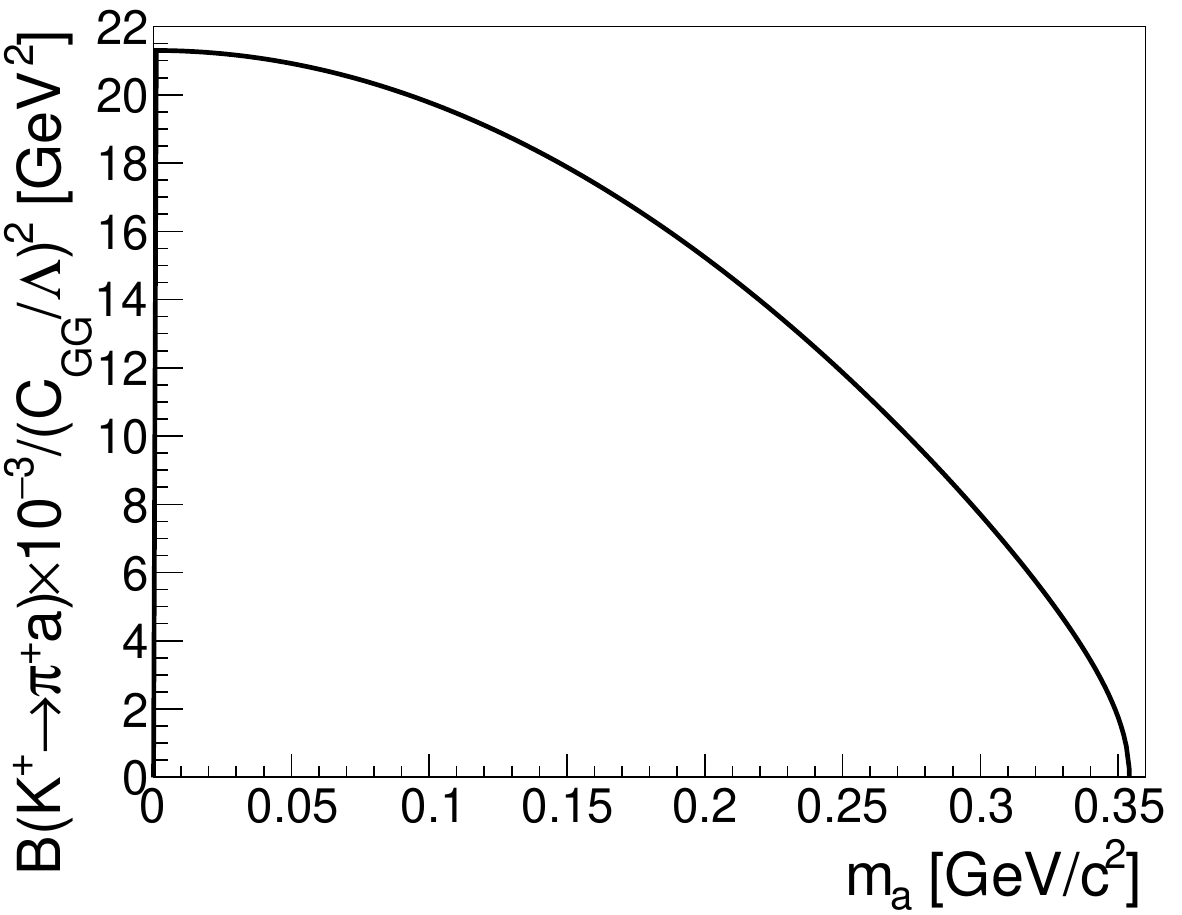}}\
\subfloat[]{\includegraphics[width=0.49\linewidth]{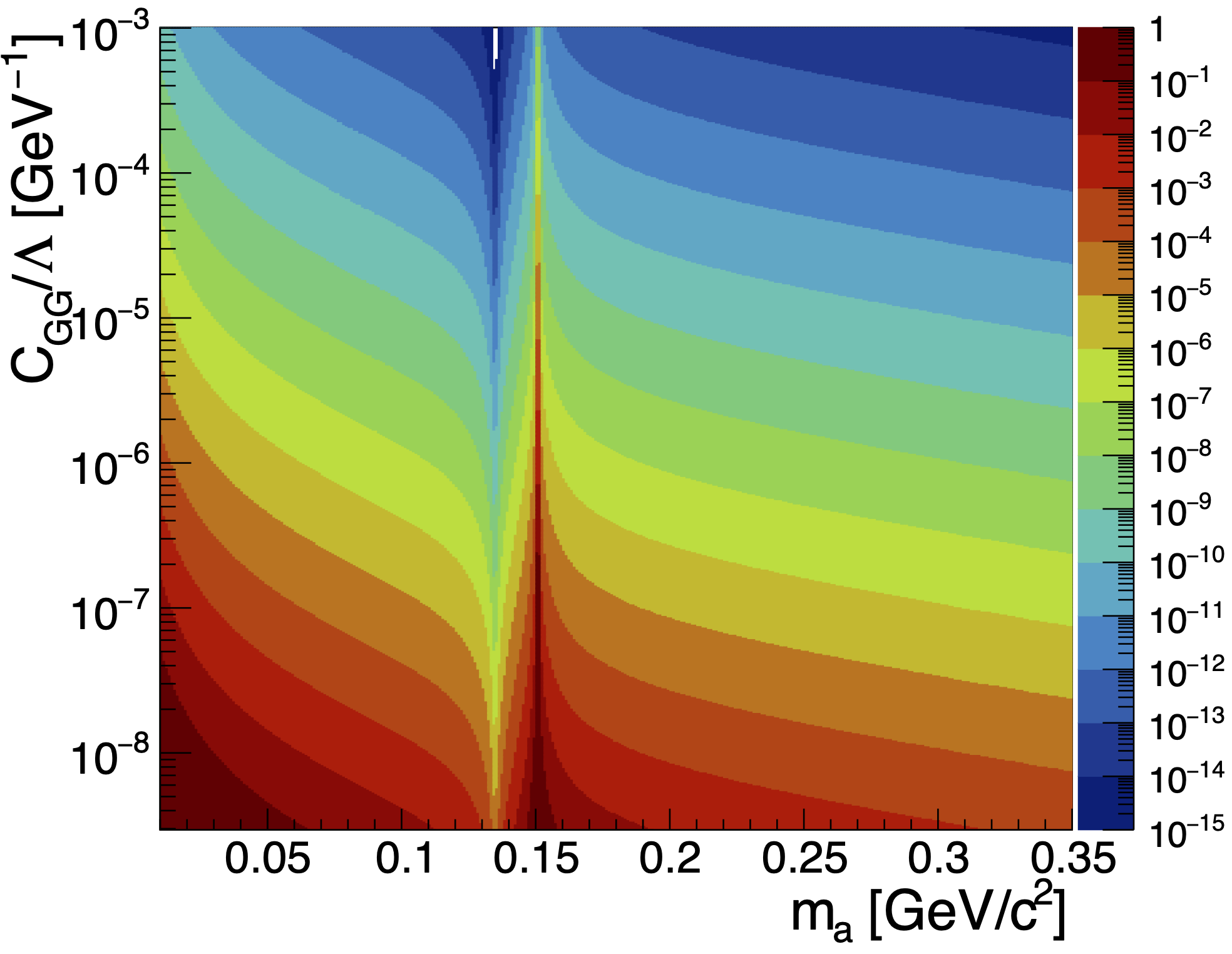}}\
\subfloat[]{\includegraphics[width=0.49\linewidth]{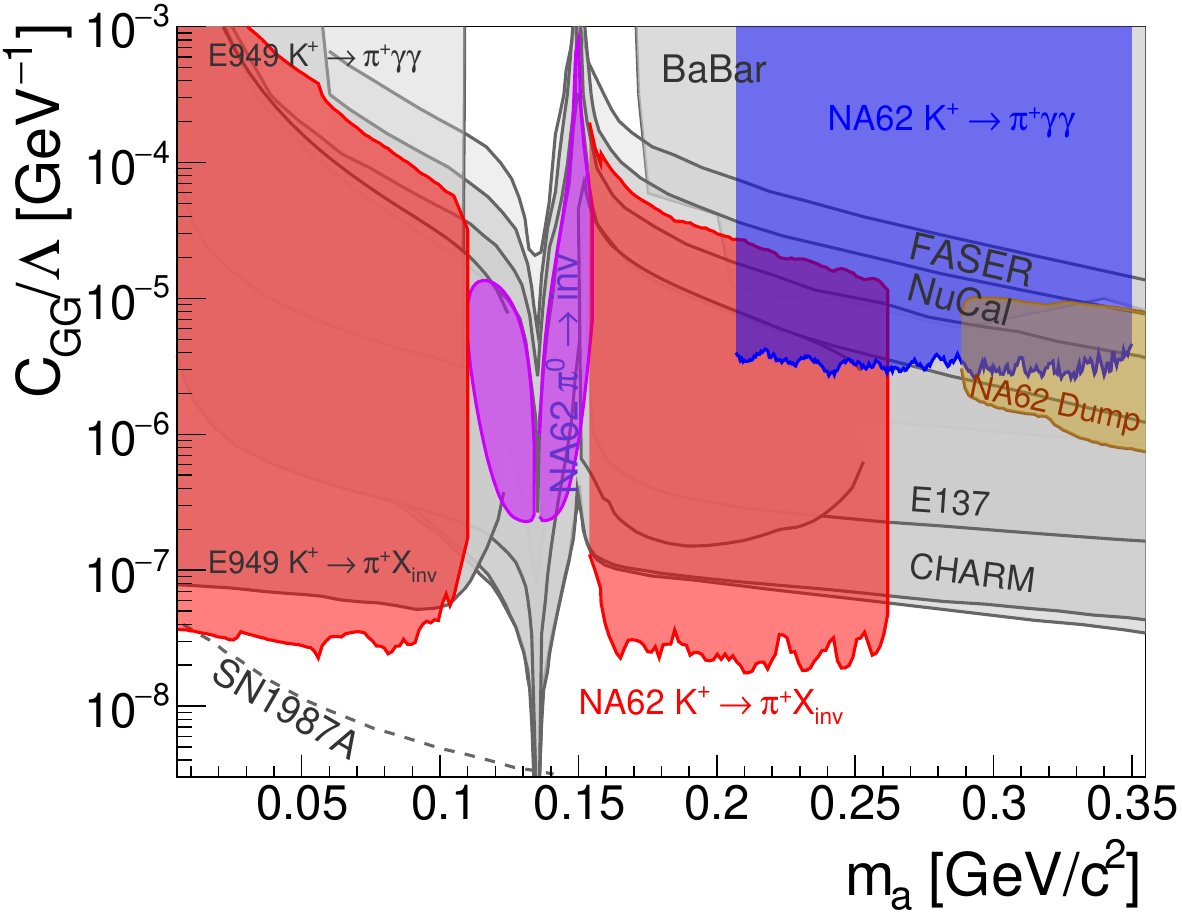}}\
\vspace{-5pt}
\caption{
(a): Branching ratio of the $K^{+}\rightarrow\pi^{+}a$ decay divided by $(C_{GG}/\Lambda)^{2}$, as a function of $m_{a}$, assuming $\Lambda = 1\,\text{TeV}$~\cite{Jerhot:2022chi}.
(b): Lifetime (colour scale, $\tau_{a}$ in s) of an ALP $a$, in the BC11 model~\cite{Beacham:2019nyx} 
as a function of mass and coupling strength~\cite{Jerhot:2022chi}.
(c): Excluded regions, at $90\,\%$ CL, of the parameter space $(m_{a},C_{GG}/\Lambda)$ for an ALP $a$, of the BC11 model~\cite{Beacham:2019nyx},
evaluated assuming $\Lambda=1\,\text{TeV}$.
Excluded regions from NA62 searches for $K^{+}\rightarrow\pi^{+}X_{\text{inv}}$ (figure~\ref{fig:KpiX_1622_limits}), 
$\pi^{0}\rightarrow\text{inv}$~\cite{NA62pi0inv},
$K^{+}\rightarrow\pi^{+}a,\,a\rightarrow\gamma\gamma$~\cite{NA62:2023olg}, 
and in beam dump mode~\cite{NA62DumpMode} are shown in red, purple, blue and brown, respectively.
Other experimental limits~\cite{Afik:2023mhj,ICARUS:2024oqb,FASER:2024bbl} are shown in grey.
}
\label{fig:ExlusionBC11}
\vspace{-5pt}
\end{figure}

In the BC11 model~\cite{Beacham:2019nyx} $a$ is an ALP with gluon coupling $C_{GG}$.
For $m_{a}<3\,m_{\pi}$, the ALP decays almost exclusively to photon pairs due to the assumption of vanishing lepton coupling. 
The branching ratio divided by $(C_{GG}/\Lambda)^{2}$ as a function of the ALP mass,
assuming $\Lambda=1\,\text{TeV}$, based on~\cite{Bauer:2021wjo} and evaluated using the \texttt{ALPINIST} framework~\cite{Jerhot:2022chi},
is displayed in figure~\ref{fig:ExlusionBC11}-a.

Figure~\ref{fig:ExlusionBC11}-b shows the ALP lifetime as a function of mass and coupling. 
Excluded regions established in the $(m_{a},C_{GG}/\Lambda)$ parameter space from the searches for $K^{+}\rightarrow\pi^{+}X_{\text{inv}}$ (figure~\ref{fig:KpiX_1622_limits}), $\pi^{0}\rightarrow\text{inv}$~\cite{NA62pi0inv} and $K^{+}\rightarrow\pi^{+}a,\,a\rightarrow\gamma\gamma$~\cite{NA62:2023olg} are displayed in figure~\ref{fig:ExlusionBC11}-c.

Within a more specific QCD axion model, constraints on the vectorial axion-down-strange coupling were established~\cite{Guadagnoli:2025xnt}, using an independent analysis procedure of the public NA62 data~\cite{NA62pubicdata_RUN1} 
from the search for the $K^{+}\rightarrow\pi^{+}X$ decay with $m_{X}=0$.

\newpage
\section{Conclusions}

Constraints on the production of a hidden-sector particle $X$ are derived from studies of the rare kaon decays $K^{+}\rightarrow\pi^{+}\nu\bar{\nu}$, $K^{+}\rightarrow\pi^{+}\mu^{+}\mu^{-}$ and $K^{+}\rightarrow\pi^{+}\gamma\gamma$.
The scenarios where $X$ is invisible or visible (decaying to a pair of SM particles) are considered.
New limits are obtained for a vector particle decaying to invisible final states, as well as for scalar and axion-like particles. 
Interpretation of the $K^+\to\pi^+\nu\bar{\nu}$ measurement to search for $K^{+}\rightarrow\pi^{+}X_{\rm inv}$, 
now extended to the 2016--2022 dataset,
provides world-leading constraints in the $m_X$ ranges $0$--$100\,\text{MeV}/c^{2}$ and $150$--$260\,\text{MeV}/c^{2}$.

\section*{Acknowledgements}
It is a pleasure to express our appreciation to the staff of the CERN laboratory and the technical
staff of the participating laboratories and universities for their efforts in the operation of the
experiment and data processing.

The cost of the experiment and its auxiliary systems was supported by the funding agencies of 
the Collaboration Institutes. We are particularly indebted to: 
F.R.S.-FNRS (Fonds de la Recherche Scientifique - FNRS), under Grants No. 4.4512.10, 1.B.258.20, Belgium;
CECI (Consortium des Equipements de Calcul Intensif), funded by the Fonds de la Recherche Scientifique de Belgique (F.R.S.-FNRS) under Grant No. 2.5020.11 and by the Walloon Region, Belgium;
NSERC (Natural Sciences and Engineering Research Council), funding SAPPJ-2018-0017,  Canada;
MEYS (Ministry of Education, Youth and Sports) funding LM 2018104, Czech Republic;
BMBF (Bundesministerium f\"{u}r Bildung und Forschung), Germany;
INFN  (Istituto Nazionale di Fisica Nucleare),  Italy;
MIUR (Ministero dell'Istruzione, dell'Universit\`a e della Ricerca),  Italy;
CONACyT  (Consejo Nacional de Ciencia y Tecnolog\'{i}a),  Mexico;
IFA (Institute of Atomic Physics) Romanian 
CERN-RO Nr. 06/03.01.2022
and Nucleus Programme PN 19 06 01 04,  Romania;
MESRS  (Ministry of Education, Science, Research and Sport), Slovakia; 
CERN (European Organization for Nuclear Research), Switzerland; 
STFC (Science and Technology Facilities Council), United Kingdom;
NSF (National Science Foundation) Award Numbers 1506088 and 1806430,  U.S.A.;
ERC (European Research Council)  ``UniversaLepto'' advanced grant 268062, ``KaonLepton'' starting grant 336581, Europe.

Individuals have received support from:
Charles University (grants UNCE 24/SCI/016, PRIMUS 23/SCI/025), 
Ministry of Education, Youth and Sports (project FORTE CZ.02.01.01/
00/22-008/0004632), Czech Republic;
Czech Science Foundation (grant 23-06770S);   
Agence Nationale de la Recherche (grant ANR-19-CE31-0009), France;
Ministero dell'Istruzione, dell'Universit\`a e della Ricerca (MIUR  ``Futuro in ricerca 2012''  grant RBFR12JF2Z, Project GAP), Italy;
Nuclemedica Soluciones, San Luis Potos\'{i}, Mexico;
the Royal Society  (grants UF100308, UF0758946), United Kingdom;
STFC (Rutherford fellowships ST/J00412X/1, ST/M005798/1), United Kingdom;
ERC (grants 268062,  336581 and  starting grant 802836 ``AxScale'');
EU Horizon 2020 (Marie Sk\l{}odowska-Curie grants 701386, 754496, 842407, 893101, 101023808).

\newpage
\bibliography{bibliography}
\newpage
\clearpage

\newcommand{\orcimg}{\raisebox{-0.3\height}{\includegraphics[height=\fontcharht\font`A]{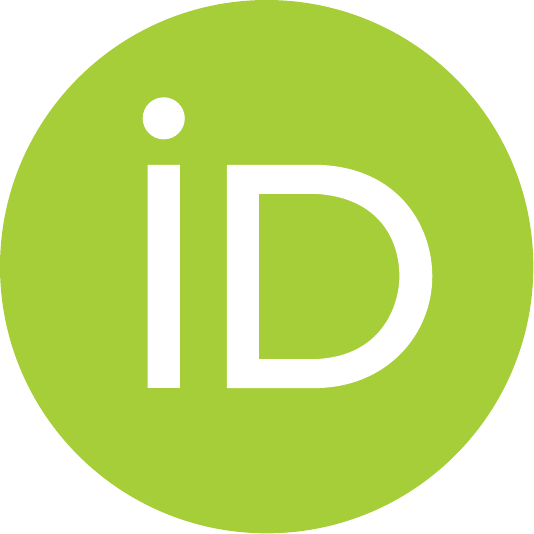}}}
\newcommand{\orcid}[1]{\href{https://orcid.org/#1}{\orcimg}}

\centerline{\bf The NA62 Collaboration}
\vspace{0.5cm}
%
%

\begin{raggedright}
\noindent
{\bf Universit\'e Catholique de Louvain, Louvain-La-Neuve, Belgium}\\
 E.~Cortina Gil\orcid{0000-0001-9627-699X},
 J.~Jerhot$\,${\footnotemark[1]}\orcid{0000-0002-3236-1471},
 E.~Minucci$\,${\footnotemark[2]}\orcid{0000-0002-3972-6824},
 S.~Padolski\orcid{0000-0002-6795-7670},
 P.~Petrov, 
 A.~Shaikhiev$\,${\footnotemark[3]}\orcid{0000-0003-2921-8743},
 R.~Volpe$\,${\footnotemark[4]}\orcid{0000-0003-1782-2978}
\vspace{0.5cm}

{\bf TRIUMF, Vancouver, British Columbia, Canada}\\
 T.~Numao\orcid{0000-0001-5232-6190},
 Y.~Petrov\orcid{0000-0003-2643-8740},
 V.~Shang\orcid{0000-0002-1436-6092},
 B.~Velghe\orcid{0000-0002-0797-8381},
 V. W. S.~Wong\orcid{0000-0001-5975-8164}
% ...
\vspace{0.5cm}

{\bf INFN, Sezione di Roma I e Dipartimento di Fisica, Sapienza Universit\`a di Roma, Roma, Italy}\\
 G.~D'Agostini\orcid{0000-0002-6245-875X},
 M.~Raggi$\,$\renewcommand{\thefootnote}{\fnsymbol{footnote}}\footnotemark[1]\renewcommand{\thefootnote}{\arabic{footnote}}\orcid{0000-0002-7448-9481}
\vspace{0.5cm}

{\bf INFN, Sezione di Roma Tor Vergata, Roma, Italy}\\
 R.~Ammendola\orcid{0000-0003-4501-3289},
 V.~Bonaiuto$\,${\footnotemark[22]}\orcid{0000-0002-2328-4793},
 A.~Fucci,
 A.~Salamon\orcid{0000-0002-8438-8983},
 F.~Sargeni$\,${\footnotemark[23]}\orcid{0000-0002-0131-236X}
\vspace{0.5cm}

{\bf INFN, Sezione di Torino, Torino, Italy}\\
 C.~Biino$\,${\footnotemark[24]}\orcid{0000-0002-1397-7246},
 A.~Filippi\orcid{0000-0003-4715-8748},
 F.~Marchetto\orcid{0000-0002-5623-8494}
\vspace{0.5cm}

{\bf INFN, Sezione di Torino e Dipartimento di Fisica dell'Universit\`a, Torino, Italy}\\
 R.~Arcidiacono$\,${\footnotemark[25]}\orcid{0000-0001-5904-142X},
 B.~Bloch-Devaux$\,${\footnotemark[6]}$^,$$\,${\footnotemark[26]}\orcid{0000-0002-2463-1232},
 E.~Menichetti\orcid{0000-0001-7143-8200},
 E.~Migliore\orcid{0000-0002-2271-5192},
 D.~Soldi\orcid{0000-0001-9059-4831}
\vspace{0.5cm}

{\bf Institute of Nuclear Physics, Almaty, Kazakhstan}\\
 Y.~Mukhamejanov\orcid{0000-0002-9064-6061},
 A.~Mukhamejanova$\,${\footnotemark[27]}\orcid{0009-0004-4799-9066},
 N.~Saduyev\orcid{0000-0002-5144-0677},
 S.~Sakhiyev\orcid{0000-0002-9014-9487}
\vspace{0.5cm}

{\bf Instituto de F\'isica, Universidad Aut\'onoma de San Luis Potos\'i, San Luis Potos\'i, Mexico}\\
 A.~Briano Olvera\orcid{0000-0001-6121-3905},
 J.~Engelfried\orcid{0000-0001-5478-0602},
 N.~Estrada-Tristan$\,${\footnotemark[28]}\orcid{0000-0003-2977-9380},
 M.~A.~Reyes~Santos$\,${\footnotemark[28]}\orcid{0000-0003-1347-2579},
 K.~A.~Rodriguez~Rivera\orcid{0000-0001-5723-9176}
\vspace{0.5cm}

{\bf Horia Hulubei National Institute for R\&D in Physics and Nuclear Engineering, Bucharest-Magurele, Romania}\\
 P.~Boboc\orcid{0000-0001-5532-4887},
 A. M.~Bragadireanu,
 S. A.~Ghinescu\orcid{0000-0003-3716-9857},
 O. E.~Hutanu
\vspace{0.5cm}

{\bf Faculty of Mathematics, Physics and Informatics, Comenius University, Bratislava, Slovakia}\\
 T.~Blazek\orcid{0000-0002-2645-0283},
 V.~Cerny\orcid{0000-0003-1998-3441},
 Z.~Kucerova$\,${\footnotemark[7]}\orcid{0000-0001-8906-3902},
 T.~Velas\orcid{0009-0004-0061-1968}
\vspace{0.5cm}

{\bf CERN, European Organization for Nuclear Research, Geneva, Switzerland}\\
 J.~Bernhard\orcid{0000-0001-9256-971X},
 L.~Bician$\,${\footnotemark[29]}\orcid{0000-0001-9318-0116},
 M.~Boretto\orcid{0000-0001-5012-4480},
 F.~Brizioli$\,$\renewcommand{\thefootnote}{\fnsymbol{footnote}}\footnotemark[1]\renewcommand{\thefootnote}{\arabic{footnote}}$^,$$\,${\footnotemark[4]}\orcid{0000-0002-2047-441X},
 A.~Ceccucci\orcid{0000-0002-9506-866X}, 
 M.~Corvino\orcid{0000-0002-2401-412X},
 H.~Danielsson\orcid{0000-0002-1016-5576},
 N.~De Simone$\,${\footnotemark[30]},
 F.~Duval,
 L.~Federici$\,${\footnotemark[31]}\orcid{0000-0002-3401-9522}, 
 E.~Gamberini\orcid{0000-0002-6040-4985},
 L.~Gatignon$\,${\footnotemark[3]}\orcid{0000-0001-6439-2945},
 R.~Guida,
 F.~Hahn$\,$\renewcommand{\thefootnote}{\fnsymbol{footnote}}\footnotemark[2]\renewcommand{\thefootnote}{\arabic{footnote}},
 E.~B.~Holzer\orcid{0000-0003-2622-6844}, 
 B.~Jenninger,
 M.~Koval$\,${\footnotemark[29]}\orcid{0000-0002-6027-317X},
 P.~Laycock$\,${\footnotemark[32]}\orcid{0000-0002-8572-5339},
 G.~Lehmann Miotto\orcid{0000-0001-9045-7853},
 P.~Lichard\orcid{0000-0003-2223-9373}, 
 A.~Mapelli\orcid{0000-0002-4128-1019},
 K.~Massri$\,${\footnotemark[3]}\orcid{0000-0001-7533-6295},
 M.~Noy,
 V.~Palladino\orcid{0000-0002-9786-9620},
 J.~Pinzino$\,${\footnotemark[33]}\orcid{0000-0002-7418-0636}, 
 V.~Ryjov,
 S.~Schuchmann$\,${\footnotemark[15]}\orcid{0000-0002-8088-4226},
 J.~Swallow$\,$\renewcommand{\thefootnote}{\fnsymbol{footnote}}\footnotemark[1]\renewcommand{\thefootnote}{\arabic{footnote}}$^,$$\,${\footnotemark[2]}\orcid{0000-0002-1521-0911}, 
S.~Venditti
\vspace{0.5cm}

{\bf Ecole Polytechnique F\'ed\'erale Lausanne, Lausanne, Switzerland}\\
 X.~Chang\orcid{0000-0002-8792-928X},
 A.~Kleimenova\orcid{0000-0002-9129-4985},
 R.~Marchevski\orcid{0000-0003-3410-0918}
\vspace{0.5cm}

{\bf School of Physics and Astronomy, University of Birmingham, Birmingham, United Kingdom}\\
 T.~Bache\orcid{0000-0003-4520-830X},
 M. B.~Brunetti$\,${\footnotemark[34]}\orcid{0000-0003-1639-3577},
 V.~Duk$\,${\footnotemark[4]}\orcid{0000-0001-6440-0087},
 V.~Fascianelli$\,${\footnotemark[35]},
 J. R.~Fry\orcid{0000-0002-3680-361X}, 
 F.~Gonnella\orcid{0000-0003-0885-1654},
 E.~Goudzovski\orcid{0000-0001-9398-4237},
 J.~Henshaw\orcid{0000-0001-7059-421X},
 L.~Iacobuzio,
 C.~Kenworthy\orcid{0009-0002-8815-0048}, 
 C.~Lazzeroni\orcid{0000-0003-4074-4787},
 N.~Lurkin\orcid{0000-0002-9440-5927},
 F.~Newson,
 C.~Parkinson\orcid{0000-0003-0344-7361},
 A.~Romano\orcid{0000-0003-1779-9122}, 
 C.~Sam\orcid{0009-0005-3802-5777},
 J.~Sanders\orcid{0000-0003-1014-094X},
 A.~Sergi$\,${\footnotemark[36]}\orcid{0000-0001-9495-6115},
 A.~Sturgess\orcid{0000-0002-8104-5571},
 A.~Tomczak\orcid{0000-0001-5635-3567}
\vspace{0.5cm}

{\bf School of Physics, University of Bristol, Bristol, United Kingdom}\\
 H.~Heath\orcid{0000-0001-6576-9740},
 R.~Page,
 S.~Trilov\orcid{0000-0003-0267-6402}
\vspace{0.5cm}

\vspace{0.9cm}
{\bf School of Physics and Astronomy, University of Glasgow, Glasgow, United Kingdom}\\
 B.~Angelucci,
 D.~Britton\orcid{0000-0001-9998-4342},
 C.~Graham\orcid{0000-0001-9121-460X},
 D.~Protopopescu\orcid{0000-0002-8047-6513}
\vspace{0.5cm}

{\bf Physics Department, University of Lancaster, Lancaster, United Kingdom}\\
 J.~Carmignani$\,${\footnotemark[37]}\orcid{0000-0002-1705-1061},
 J. B.~Dainton,
 R. W. L.~Jones\orcid{0000-0002-6427-3513},
 G.~Ruggiero$\,${\footnotemark[38]}\orcid{0000-0001-6605-4739}
\vspace{0.5cm}

{\bf School of Physical Sciences, University of Liverpool, Liverpool, United Kingdom}\\
 L.~Fulton,
 D.~Hutchcroft\orcid{0000-0002-4174-6509},
 E.~Maurice$\,${\footnotemark[39]}\orcid{0000-0002-7366-4364},
 B.~Wrona\orcid{0000-0002-1555-0262}
\vspace{0.5cm}

{\bf Physics and Astronomy Department, George Mason University, Fairfax, Virginia, USA}\\
 A.~Conovaloff,
 P.~Cooper,
 D.~Coward$\,${\footnotemark[40]}\orcid{0000-0001-7588-1779},
 P.~Rubin\orcid{0000-0001-6678-4985}
\vspace{0.5cm}

{\bf Authors affiliated with an Institute formerly covered by a cooperation agreement with CERN}\\
 S.~Fedotov,
 E.~Gushchin\orcid{0000-0001-8857-1665},
 S.~Kholodenko$\,${\footnotemark[33]}\orcid{0000-0002-0260-6570},
 A.~Khotyantsev,
 Y.~Kudenko\orcid{0000-0003-3204-9426}, 
 V.~Kurochka,
 V.~Kurshetsov\orcid{0000-0003-0174-7336},
 M.~Medvedeva,
 A.~Mefodev,
 V.~Obraztsov\orcid{0000-0002-0994-3641}, 
 A.~Ostankov$\,$\renewcommand{\thefootnote}{\fnsymbol{footnote}}\footnotemark[2]\renewcommand{\thefootnote}{\arabic{footnote}},
 V.~Semenov$\,$\renewcommand{\thefootnote}{\fnsymbol{footnote}}\footnotemark[2]\renewcommand{\thefootnote}{\arabic{footnote}},
 V.~Sugonyaev\orcid{0000-0003-4449-9993},
 O.~Yushchenko\orcid{0000-0003-4236-5115}
\vspace{0.5cm}

{\bf Authors affiliated with an international laboratory covered by a cooperation agreement with CERN}\\
 A.~Baeva,
 D.~Baigarashev$\,${\footnotemark[41]}\orcid{0000-0001-6101-317X},
 V.~Bautin\orcid{0000-0002-5283-6059},
 D.~Emelyanov,
 T.~Enik\orcid{0000-0002-2761-9730}, 
 V.~Falaleev$\,${\footnotemark[4]}\orcid{0000-0003-3150-2196},
 K.~Gorshanov\orcid{0000-0001-7912-5962},
 V.~Kekelidze\orcid{0000-0001-8122-5065},
 D.~Kereibay,
 A.~Korotkova, 
 L.~Litov$\,${\footnotemark[17]}\orcid{0000-0002-8511-6883},
 D.~Madigozhin\orcid{0000-0001-8524-3455},
 M.~Misheva$\,${\footnotemark[42]},
 N.~Molokanova,
 S.~Movchan, 
 A.~Okhotnikov\orcid{0000-0003-1404-3522},
 I.~Polenkevich,
 Yu.~Potrebenikov\orcid{0000-0003-1437-4129},
 A.~Sadovskiy\orcid{0000-0002-4448-6845},
 K.~Salamatin\orcid{0000-0001-6287-8685}, 
 S.~Shkarovskiy,
 A.~Zinchenko$\,$\renewcommand{\thefootnote}{\fnsymbol{footnote}}\footnotemark[2]\renewcommand{\thefootnote}{\arabic{footnote}}
\vspace{0.5cm}

\end{raggedright}
\vspace{0.5cm}
%
%

\setcounter{footnote}{0}
\newlength{\basefootnotesep}
\setlength{\basefootnotesep}{\footnotesep}

\renewcommand{\thefootnote}{\fnsymbol{footnote}}
\noindent
$^{\footnotemark[1]}${Corresponding authors: F.~ Brizioli, M.~Raggi, J.~Swallow,  \\
email: francesco.brizioli@cern.ch, mauro.raggi@cern.ch, joel.christopher.swallow@cern.ch}\\
$^{\footnotemark[2]}${Deceased}\\
\renewcommand{\thefootnote}{\arabic{footnote}}
$^{1}${Present address: Max-Planck-Institut f\"ur Physik (Werner-Heisenberg-Institut), Garching, D-85748, Germany} \\
$^{2}${Present address: INFN, Laboratori Nazionali di Frascati, I-00044 Frascati, Italy} \\
$^{3}${Present address: Physics Department, University of Lancaster, Lancaster, LA1 4YB, UK} \\
$^{4}${Present address: INFN, Sezione di Perugia, I-06100 Perugia, Italy} \\
$^{5}${Also at TRIUMF, Vancouver, British Columbia, V6T 2A3, Canada} \\
$^{6}${Also at School of Physics and Astronomy, University of Birmingham, Birmingham, B15 2TT, UK} \\
$^{7}${Present address: CERN, European Organization for Nuclear Research, CH-1211 Geneva 23, Switzerland} \\
$^{8}${Also at Universit\'e de Toulon, Aix Marseille University, CNRS, IM2NP, F-83957, La Garde, France} \\
$^{9}${Also at Department of Physics, Technical University of Munich, M\"unchen, D-80333, Germany} \\
$^{10}${Present address: Institut f\"ur Kernphysik and Helmholtz Institute Mainz, Universit\"at Mainz, Mainz, D-55099, Germany} \\
$^{11}${Also at CERN, European Organization for Nuclear Research, CH-1211 Geneva 23, Switzerland} \\
$^{12}${Present address: Universit\"at W\"urzburg, D-97070 W\"urzburg, Germany} \\
$^{13}${Present address: European XFEL GmbH, D-22869 Schenefeld, Germany} \\
$^{14}${Present address: School of Physics and Astronomy, University of Glasgow, Glasgow, G12 8QQ, UK} \\
$^{15}${Present address: Institut f\"ur Physik and PRISMA Cluster of Excellence, Universit\"at Mainz, D-55099 Mainz, Germany} \\
$^{16}${Also at Dipartimento di Scienze Fisiche, Informatiche e Matematiche, Universit\`a di Modena e Reggio Emilia, I-41125 Modena, Italy} \\
$^{17}${Also at Faculty of Physics, University of Sofia, BG-1164 Sofia, Bulgaria} \\
$^{18}${Also at INFN, Sezione di Roma I e Dipartimento di Fisica, Sapienza Universit\`a di Roma, I-00185 Roma, Italy} \\
$^{19}${Present address: INFN, Sezione di Napoli e Scuola Superiore Meridionale, I-80138 Napoli, Italy} \\
$^{20}${Present address: Instituto de F\'isica, Universidad Aut\'onoma de San Luis Potos\'i, 78240 San Luis Potos\'i, Mexico} \\
$^{21}${Present address: INFN, Sezione di Roma I, I-00185 Roma, Italy} \\
$^{22}${Also at Department of Industrial Engineering, University of Roma Tor Vergata, I-00173 Roma, Italy} \\
$^{23}${Also at Department of Electronic Engineering, University of Roma Tor Vergata, I-00173 Roma, Italy} \\
$^{24}${Also at Gran Sasso Science Institute, I-67100 L'Aquila,  Italy} \\
$^{25}${Also at Universit\`a degli Studi del Piemonte Orientale, I-13100 Vercelli, Italy} \\
$^{26}${Present address: Universit\'e Catholique de Louvain, B-1348 Louvain-La-Neuve, Belgium} \\
$^{27}${Also at al-Farabi Kazakh National University, 050040 Almaty, Kazakhstan} \\
$^{28}${Also at Universidad de Guanajuato, 36000 Guanajuato, Mexico} \\
$^{29}${Present address: Charles University, 116 36 Prague 1, Czech Republic} \\
$^{30}${Present address: DESY, D-15738 Zeuthen, Germany} \\
$^{31}${Present address: IPHC, CNRS/IN2P3, Strasbourg University, F-67037 Strasbourg, France} \\
$^{32}${Present address: Brookhaven National Laboratory, Upton, NY 11973, USA} \\
$^{33}${Present address: INFN, Sezione di Pisa, I-56100 Pisa, Italy} \\
$^{34}${Present address: Kansas University, Lawrence, KS 66045-7582, USA} \\
$^{35}${Present address: Center for theoretical neuroscience, Columbia University, New York, NY 10027, USA} \\
$^{36}${Present address: INFN, Sezione di Genova e Dipartimento di Fisica dell'Universit\`a, I-16146 Genova, Italy} \\
$^{37}${Present address: School of Physical Sciences, University of Liverpool, Liverpool, L69 7ZE, UK} \\
$^{38}${Present address: INFN, Sezione di Firenze e Dipartimento di Fisica e Astronomia dell'Universit\`a, I-50019 Sesto Fiorentino, Italy} \\
$^{39}${Present address: Laboratoire Leprince Ringuet, F-91120 Palaiseau, France} \\
$^{40}${Also at SLAC National Accelerator Laboratory, Stanford University, Menlo Park, CA 94025, USA} \\
$^{41}${Also at L. N. Gumilyov Eurasian National University, 010000 Nur-Sultan, Kazakhstan} \\
$^{42}${Present address: Institute of Nuclear Research and Nuclear Energy of Bulgarian Academy of Science (INRNE-BAS), BG-1784 Sofia, Bulgaria} \\

\end{document}